\documentclass[a4paper,11pt]{article}
\pdfoutput=1 

\usepackage{jheppub} 

\usepackage[T1]{fontenc} 
\usepackage[multiple]{footmisc}
\usepackage{snapshot}

\preprint{\parbox{3cm}{ZU-TH 30/13\\MCnet-13-19\\LPN13-120}}

\title{Higgs Boson pair production merged to one jet}

\author[1]{Philipp Maierh\"ofer,}
\author[2]{Andreas Papaefstathiou,}

\affiliation[1,2]{Institut f\"ur Theoretische Physik, Universit\"at Z\"urich, Switzerland}
\emailAdd{andreasp@physik.uzh.ch}
\emailAdd{philipp@physik.uzh.ch}

\abstract{We develop a Monte Carlo event generator for Higgs Boson pair 
  production merged to exact one-jet matrix elements. The
  matrix elements are generated with \texttt{OpenLoops} and event
  generation is performed with the \texttt{HERWIG++} general-purpose
  event generator. This allows us to simulate fully-exclusive hadronic
  final states with accurate description of the kinematics of the
  leading jet in conjunction with a parton shower. We use the implementation to examine in detail the systematic
  uncertainties which result from the merging procedure. We assess the
  magnitude of the impact of the merging on experimental
  searches of Standard Model di-Higgs production that aim to constrain
  the Higgs boson self-coupling. We find that the use of a merged
  sample can reduce theoretical systematic uncertainties in the
  efficiencies of cuts on certain observables. This constitutes the most
  accurate simulation of the process available to date. The Monte Carlo event generator developed for this project is
  available as an add-on to the \texttt{HERWIG++} event generator at \url{http://www.physik.uzh.ch/data/openloops/download/projects/hhmerge/}.}

\begin{document}

\maketitle
\flushbottom

\section{Introduction}

\subsection{Context and introductory remarks}

The Large Hadron Collider ATLAS and CMS experiments confirmed the
existence of a scalar particle consistent with the Standard
Model (SM) Higgs Boson~\cite{ATLAS_Higgs,
  CMS_Higgs, CMS-PAS-HIG-12-045, ATLAS:2012wma,
  ATLAS-CONF-2012-170}. However the quest for understanding the
mechanism behind the electroweak symmetry breaking (EWSB) does not end
with the discovery of the Higgs Boson; measuring its couplings to the SM fields is an important and long-term task that
the ATLAS and CMS experiments, as well as future collider experiments,
are expected to undertake. It is crucial, moreover, to determine
whether the realisation of the mechanism of the EWSB is indeed
SM-like. This can be investigated by examining the Higgs potential
which, after EWSB in the minimal prescription, can be written as
\begin{equation}
\mathcal{V}(h) = \frac{1}{2} m^2_h h^2 + \lambda_{hhh} v h^3 +
\frac{1}{4} \lambda_{hhhh} h^4\;.
\end{equation}
Within the SM we have $\lambda^{\mathrm{SM}}_{hhh} =
\lambda^{\mathrm{SM}}_{hhhh} = m_h^2 / (2v^2) \simeq 0.13$ for a Higgs
boson mass of $m_h \simeq 125$~GeV. The discovery of the Higgs boson only
indicates the size of the curvature of the potential around the local
minimum, coming from the quadratic term. To confirm the form of the potential, the
measurement of higher-order terms is necessary. At the LHC, these terms can be probed directly via double or triple Higgs Boson
production. The tiny cross section for triple Higgs production makes it impossible to perform any meaningful measurement in the foreseeable
future, even during the full life-time of the LHC~\cite{Plehn:2005nk, Binoth:2006ym}. Higgs Boson pair production, on the other hand, is
certainly challenging but not impossible to observe at the LHC. Interesting phenomenological
studies were performed more than 10 years ago~\cite{Glover:1987nx,
  Dawson:1998py, Djouadi:1999rca, Plehn:1996wb, Baur:2002qd,
  Baur:2003gp} and more
recently, owing to the discovery of the Higgs Boson as well as the
development of boosted jet techniques, the subject has undergone a
lively rejuvenation~\cite{Baglio:2012np, Barr:2013tda, Dolan:2013rja,
  Papaefstathiou:2012qe, Goertz:2013kp,
  Goertz:2013eka, deFlorian:2013jea, deFlorian:2013uza, Grigo:2013rya,
  Cao:2013si, Gupta:2013zza, Nhung:2013lpa, Ellwanger:2013ova, No:2013wsa, McCullough:2013rea}. 

Despite the fact that several interesting and in-depth
phenomenological studies of inclusive Higgs Boson pair production at
the LHC ($pp
\rightarrow hh + X$) have been performed, the Monte Carlo event simulation of the process has relied so far only on leading-order matrix
elements with the addition of parton showers to simulate the extra QCD
radiation.\footnote{During the final stages of preparation of this
  article, a similar study has appeared in
  Ref.~\cite{Li:2013flc}. Here we provide a completely independent
  implementation, both in terms of the merging and the production frameworks employed.} Exceptions to this are two recent studies which examined the
exclusive one- and two-jet channels in the full theory, with the full
top mass dependence, (i.e.\ $pp \rightarrow hhj + X$ and
$pp \rightarrow hhjj +X$) and contrasted these to results obtained in the effective
theory~\cite{Dolan:2012rv, Dolan:2013rja}. It is important to stress, however, that the kinematical
properties of inclusive final states can be substantially altered by
the inclusion of higher-order matrix elements. This is especially true
in the inclusive $hh+X$ process, which is predominantly gluon-gluon initiated,
and hence is inevitably accompanied by a copious amount of QCD
radiation. Thus, the accuracy, and hence reliability, of the
kinematics of inclusive di-Higgs searches will certainly benefit from
the inclusion of the exact real-emission higher-order matrix elements.

\subsection{Di-Higgs production at higher orders}\label{sec:hh-ho}

\begin{figure}[!htb]
  \begin{center}
    \vspace*{1ex}
    \includegraphics{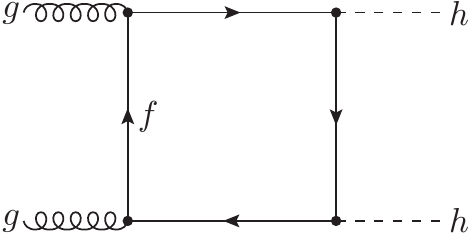}\qquad
    \includegraphics{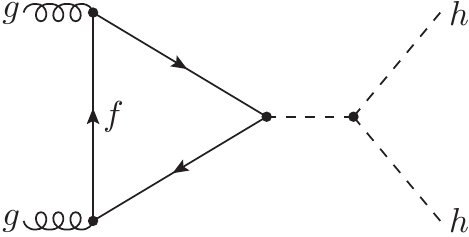}
    \vspace{-2ex}
  \end{center}
  \caption{The Higgs pair production diagrams contributing to the
    gluon fusion process at LO are shown for a generic fermion $f$.}
  \label{fig:HHdiagrams}
\end{figure}

The $pp \rightarrow hh+X$ process at hadron colliders is loop-induced
at leading order, proceeding via a heavy quark loop. The leading-order gluon fusion diagrams are shown in
Fig.~\ref{fig:HHdiagrams}. Evidently, a next-to-leading order calculation would
involve, among others, diagrams with two loops that
involve heavy fermions, and hence two mass scales (the fermion mass
and the Higgs Boson mass). Such diagrams currently lie at the
frontier of higher-order loop calculations. Consequently, this impedes
the implementation of a matched next-to-leading order (NLO) plus shower simulation.

The effective theory approximation in the heavy top mass limit that has been employed
in single Higgs boson production has been shown to
be insufficient to describe the kinematics of the $hh$ process, both at
leading order~\cite{Dawson:2012mk, Grigo:2013rya}, and at higher
orders~\cite{Dolan:2012rv, Dolan:2013rja}. Nevertheless, inclusive
NLO~\cite{Dawson:1998py, Dawson:1998xu} and NNLO cross
section calculations~\cite{deFlorian:2013jea, deFlorian:2013uza} have been
performed in the effective theory approximation, giving an estimate of
the size of the higher-order corrections in the full theory. 

Thus, in the absence of the full NLO calculation, the best one can do is merge samples of
different jet multiplicities in a consistent way, carefully avoiding
any issues that may arise due to double-counting or phase space region
mismatch. Such simulations have
been shown to reliably describe the kinematical properties of
experimental data (see for example the relevant experimental CMS~\cite{Chatrchyan:2011ne} and ATLAS~\cite{Aad:2012en} analyses), modulo the correct normalisation taken from
higher-order cross section calculations. Here we perform such a
merging of samples, including the full top and bottom mass dependence in
the fermion loops of Fig.~\ref{fig:HHdiagrams}, as well as the higher-order real
emission diagrams which we examine below.

This paper is organised in the following way: in
Section~\ref{sec:openloops} we briefly describe the \texttt{OpenLoops} generator
for one-loop matrix elements and provide cross sections for the various
contributing exclusive channels. In Section~\ref{sec:results} we present
results and examine the systematic uncertainties associated with the
merging prescription, and in Section~\ref{sec:pheno} we investigate
the phenomenological implications of including the merged
higher-order matrix elements. We present our conclusions in Section~\ref{sec:conclusions}.

\section{OpenLoops and matrix elements}\label{sec:openloops}

\subsection{The OpenLoops matrix element generator}

The \texttt{OpenLoops} generator is based on the open-loops algorithm~\cite{Cascioli:2011va} for the efficient evaluation of one-loop matrix elements. The algorithm employs a numerical recursion to construct the loop momentum dependence of the numerator of loop amplitudes combined with tensor integral reduction. The tensor integrals are computed by the \texttt{Collier} library, which implements the Denner-Dittmaier reduction procedure for the numerically stable evaluation of tensor integrals~\cite{Denner:2002ii,Denner:2005nn} and the scalar integrals of Ref.~\cite{Denner:2010tr}.

Incidentally, using tensor integrals allows for a high degree of optimisation through caching, since the integrals can be shared across different Feynman diagrams for all helicity and colour configurations. For on-shell reduction approaches this is only possible when the loop amplitude is interfered with a tree amplitude~\cite{Cascioli:2011va} and therefore not in calculations of loop induced processes, like the one presented here.

\subsection{Higgs pair production matrix elements}

Like in the case of single Higgs
boson production, the $hh$ production cross section at hadron colliders is
dominated by the gluon fusion channels. For a more detailed
dissection of the leading order cross section, we refer the reader to
Section~2 of Ref.~\cite{Goertz:2013kp}. 

\begin{figure}[!htb]
  \begin{center}
    \vspace*{1ex}
    \includegraphics[width=0.28\linewidth]{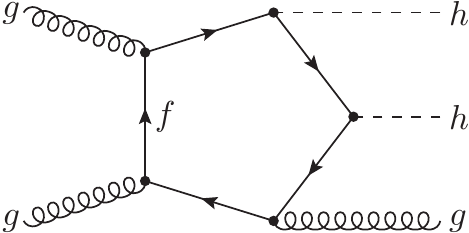}\qquad
    \includegraphics[width=0.28\linewidth]{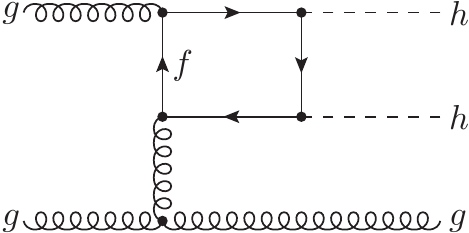}\qquad
    \includegraphics[width=0.28\linewidth]{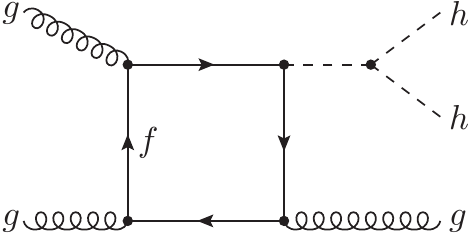}\\[3ex]
    \includegraphics[width=0.28\linewidth]{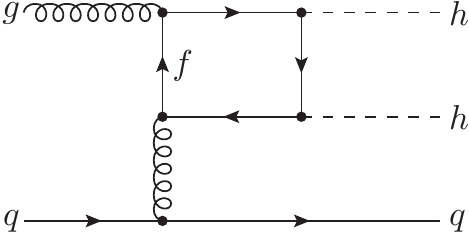}\qquad
    \includegraphics[width=0.28\linewidth]{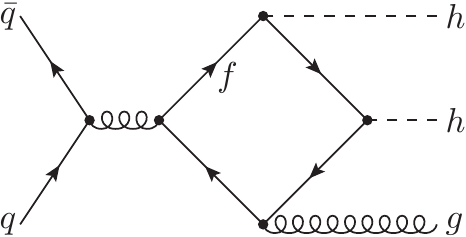}\qquad
    \includegraphics[width=0.28\linewidth]{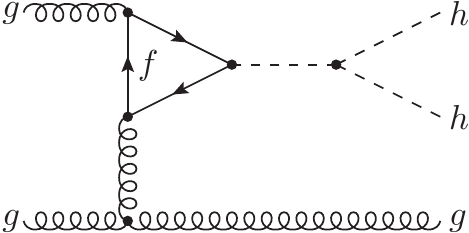}\\[3ex]
    \includegraphics[width=0.28\linewidth]{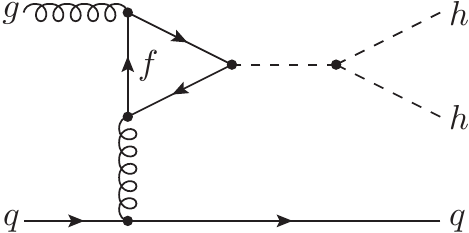}\qquad
    \includegraphics[width=0.28\linewidth]{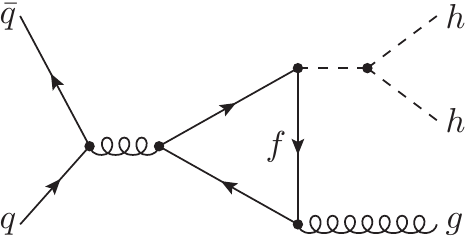}
    \vspace{-2ex}
  \end{center}
  \caption{Diagram classes which contribute to Higgs
    boson pair production in association with one extra parton are shown for a generic fermion
    $f$ running in the loop.}
  \label{fig:HHjdiagrams}
\end{figure}

The classes of higher-order real emission diagrams that we include
in our calculation are shown in Fig.~\ref{fig:HHjdiagrams}. The real
emission process also contains diagrams with $qg$, $\bar{q}g$ and
$q\bar{q}$ initial states which are subdominant but
non-negligible and must be included for consistent merging since the
parton shower introduces $g\to q\bar{q}$ splittings on the initial state gluons of the 0-jet matrix elements~\cite{Cascioli:2013gfa}. It is important to
note that the diagrams that involve radiation from
the heavy quark loop, are not included in any limit
in the parton shower Monte Carlos with which we merge
with. Hence, an intrinsic assumption of the merging procedure is
that these diagrams are sub-dominant with respect to the initial state radiation
in the parton shower-dominated regime. 

The \texttt{OpenLoops} process libraries to compute matrix elements have been
interfaced with \texttt{HERWIG++}. These can be used stand-alone, i.e.\ without the
merging, to perform studies of leading-order $hh$ production, or
$hh+j$ production. In Table~\ref{tb:hhjxsect} we present the cross sections for the different sub-processes
contributing to $pp \rightarrow hhj + X$, where $j$ is an associated
parton.\footnote{The \texttt{HERWIG++}
  implementation has been crossed-check against the
  \texttt{SHERPA} event generator~\cite{Gleisberg:2008ta}.} Here, and throughout this paper, we use the
4-flavour MSTW2008nlo 68\% confidence level parton density
functions~\cite{Martin:2009iq,Martin:2009bu,Martin:2010db}. Obviously, even with
the relatively high $p_\perp$ cut of 60~GeV, the real emission
sub-processes possess a cross section that is comparable to the leading
order gluon fusion process. This is an indication that they are indeed
significant and have to be considered for an accurate description of
the kinematics, even in an inclusive $hh+X$ analysis.

\begin{table}[t!]
\small
  \begin{center}
    \begin{tabular}{|c|c|c|c|c|c|}
      \hline
      Process & $gg\rightarrow hh$ & $gg\rightarrow hhg$ & $gq
      \rightarrow hhq$ &$g\bar{q} \rightarrow hh\bar{q}$ & $q\bar{q}
      \rightarrow hhg$ \\ \hline
      $\sigma (14~\mathrm{TeV})$ [fb] & 26.2(1) & 9.5(1) & 1.80(2) &
      0.411(6) & 0.062(1) \\ \hline
      $\sigma (33~\mathrm{TeV})$ [fb] & 145(3) & 70.2(9) & 10.0(1) & 3.39(5) &  0.206(3)\\ \hline
      $\sigma (100~\mathrm{TeV})$ [fb] & 883(5) & 555(7) & 60.6(9) & 27.1(4) & 0.79(1) \\ \hline
    \end{tabular}
  \end{center}
  \caption{Cross sections for the partonic $pp \rightarrow hh + X$ and
    for the sub-processes contributing to $pp\rightarrow hhj + X$ at
    14, 33 and 100~TeV. For the case of real emission, a cut of $p_\perp >
  60$~GeV was placed on the associated parton. The
  factorisation/renormalisation scales were both fixed to $\mu = m_h +
  p_{\perp,j}$, where $p_{\perp,j}$ is the transverse momentum of the
  associated parton in the centre of mass frame.}
\label{tb:hhjxsect}
\end{table}

\section{Merging}\label{sec:results}

\subsection{Merging methods}

In order to obtain a realistic simulation of processes involving associated
high-$p_{\scriptscriptstyle T}$ jet production, e.g. W/Z/Higgs+jets, the parton
shower approximation for the generation of soft and collinear QCD radiation must
be supplemented by high multiplicity leading-order matrix elements. Matrix
element-parton shower merging schemes, such as the so-called MLM~\cite{Mangano:2001xp,Mangano:2004lu,Alwall:2007fs} and
CKKW~\cite{Catani:2001cc,Mangano:2001xp,Mangano:2004lu,Lonnblad:2001iq,Krauss:2002up,Mrenna:2003if} methods,
have been developed for this purpose. These methods work by partitioning phase
space, by means of a jet algorithm, such that the distribution of jets
corresponds to that of the partons in the matrix elements, while the distribution
of radiation inside the jets is appropriately developed by the shower. In addition,
both the MLM and CKKW algorithms augment the distribution of radiation in the matrix
element region with Sudakov suppression effects, not present in the matrix elements
themselves, thus smoothing the transition from one radiation pattern to another
at the phase space partition.\footnote{For a full, comparative description of the
available schemes, see Ref.~\cite{Alwall:2007fs}.}\footnote{It is also
conceivable, at least in the case of one extra associated parton, to
perform a simulation with the MC@NLO or POWHEG matching prescriptions, with an
arbitrary virtual contribution which can be set to zero~\cite{Nason:2004rx, Frixione:2007vw, Frixione:2010wd, Frixione:2002ik}.}

\texttt{HERWIG++}~\cite{Gieseke:2011na, Arnold:2012fq, Bahr:2008pv, Bellm:2013lba} includes an implementation of the MLM merging
scheme. The current version of the
merging algorithm has been validated against its {\tt FORTRAN} \texttt{HERWIG}~\cite{Corcella:2000bw} counterpart
for several processes. For the purposes of this project, the
implementation has undergone minor modifications, to accommodate the
use of internally-generated matrix elements. We use this algorithm in conjunction with
the parton shower in order to merge the two. We fix the
factorisation and renormalization scales to be equal, $\mu_F =
\mu_R = \mu = \nu (m_{h} + p^{hh}_\perp)$, where $\nu$ is a parameter which we
vary, $m_{h}$ and $p^{hh}_\perp$ are the Higgs boson mass and the transverse momentum (as defined in the centre-of-mass
frame of the hard process) of the Higgs boson pair respectively. Note
that for the LO $hh$ process, $p^{hh}_\perp = 0$ and hence this implies
that $\mu = \nu m_h $ for all, even showered, LO samples. We call the merging scale $E_{Tclus}$, inspired by the way the MLM method is
implemented in the \texttt{HERWIG++} generator. We call the lowest-order sample `0-jet' and the
sample including one real emission `1-jet'. Broadly speaking, after
showering is performed in \texttt{HERWIG++}, the MLM method will effectively veto all events in the `0-jet' sample that
contain a jet with transverse momentum larger than $E_{Tclus}$. This
will result in what we will call the `0-jet exclusive' sample. In the
showered `1-jet' sample the MLM algorithm will effectively veto any
events with jets that have not `matched'\footnote{The term `matched'
  in the MLM prescription
  refers to whether a jet is found to be within a certain distance $\Delta R$,
  from a given hard parton that appears in the pre-showered event. By
  default this is taken to be $1.5\times
  R_{clus}$, where the $R_{clus}$ is the clustering cone size used in
  the merging. } the given extra parton
produced in association with the Higgs boson pair, as well as events that contain
jets harder than the `matched' jet. The resulting sample is called
`1-jet inclusive', meaning it contains no 0-jet contributions but
contains jets coming from the shower, of lower $p_\perp$ than the matrix
element parton. For more details on the algorithm
see, for example,~\cite{Alwall:2007fs}. 

Recently it has been shown~\cite{Frederix:2012ps} that the merging of samples of different multiplicities may result in
kinks due to the presence of a significant mismatch in the description of extra emissions
between the parton shower and the matrix element calculations in the
region chosen for merging. In Ref.~\cite{Frederix:2012ps}, it was
suggested that one can obtain smooth matching by the use of a smoothing
`$D$-function' that contains two scales instead of a single
scale. Sudakov reweighting was also used to achieve an even smoother
matching. Here we perform a variant of the former method: we
generate a merging scale randomly in a given interval according to a given distribution. This merging `range' is then characterised by two scales,
an `average' merging scale $\bar{E}_{Tclus}$ and a `variation' scale
$\epsilon_{clus}$. The merging scale is then randomly chosen on an
event-by-event basis using two different `schemes'. The first scheme
uses a Sine function (`Sinusoidal'),
\begin{equation}
E_{Tclus} = \frac{2 \epsilon_{clus}}{\pi} \sin^{-1} (2x -1) + \bar{E}_{Tclus}
\;\;,
\end{equation}
and the second a linear function (`Uniform'),
\begin{equation}
E_{Tclus} = (2x - 1) \epsilon_{clus} + \bar{E}_{Tclus}
\;\;,
\end{equation}
where $x \in [0,1]$ is a uniform random number. The effect of these
schemes is to smooth out the unphysical discontinuities, resulting in a continuous
merging of the shower and matrix element descriptions. 

To further improve the merging between the two samples, we perform `$\alpha_S$-reweighting' of the 1-jet matrix elements according to (schematically)
\begin{equation}
|\mathcal{M}|^2 \rightarrow |\mathcal{M}|^2  \frac{\alpha_S [ (p^{hh}_\perp)^2 ]}{\alpha_S (\mu^2)}\;,
\end{equation}
where $p^{hh}_\perp$ is the transverse momentum of the Higgs boson pair as defined in the centre-of-mass
frame of the hard process (or the transverse momentum of the
associated extra parton) and $\mu$ is the
renormalization scale. This is to
accommodate the difference between the scale that the shower uses in the
calculation of the strong coupling constant $\alpha_S$ wrt.\ the scale used in the matrix elements. In practice
the effect of the reweighting is small, especially in comparison to
the uncertainties arising from variations of $\mu$ and the merging
scale parameters. All of the results in the rest of the paper include
$\alpha_S$-reweighting.

\subsection{Systematic uncertainties}

In what follows we present results obtained at parton level, using the 
\texttt{Rivet} analysis framework version 1.8.3~\cite{Buckley:2010ar} and the anti-$k_T$ algorithm~\cite{Cacciari:2008gp} with $R=0.4$. In all the calculations, the Higgs boson mass was chosen to
be $m_h = 125$~GeV and the top quark and bottom quark masses to be
174.2~GeV and 4.7~GeV respectively, with all the widths set to
zero. The renormalization and factorisation scales were both set to
equal $\mu$: $\mu_F = \mu_R = \mu$. 

\begin{figure}
\centering
    \includegraphics[width=0.40\linewidth]{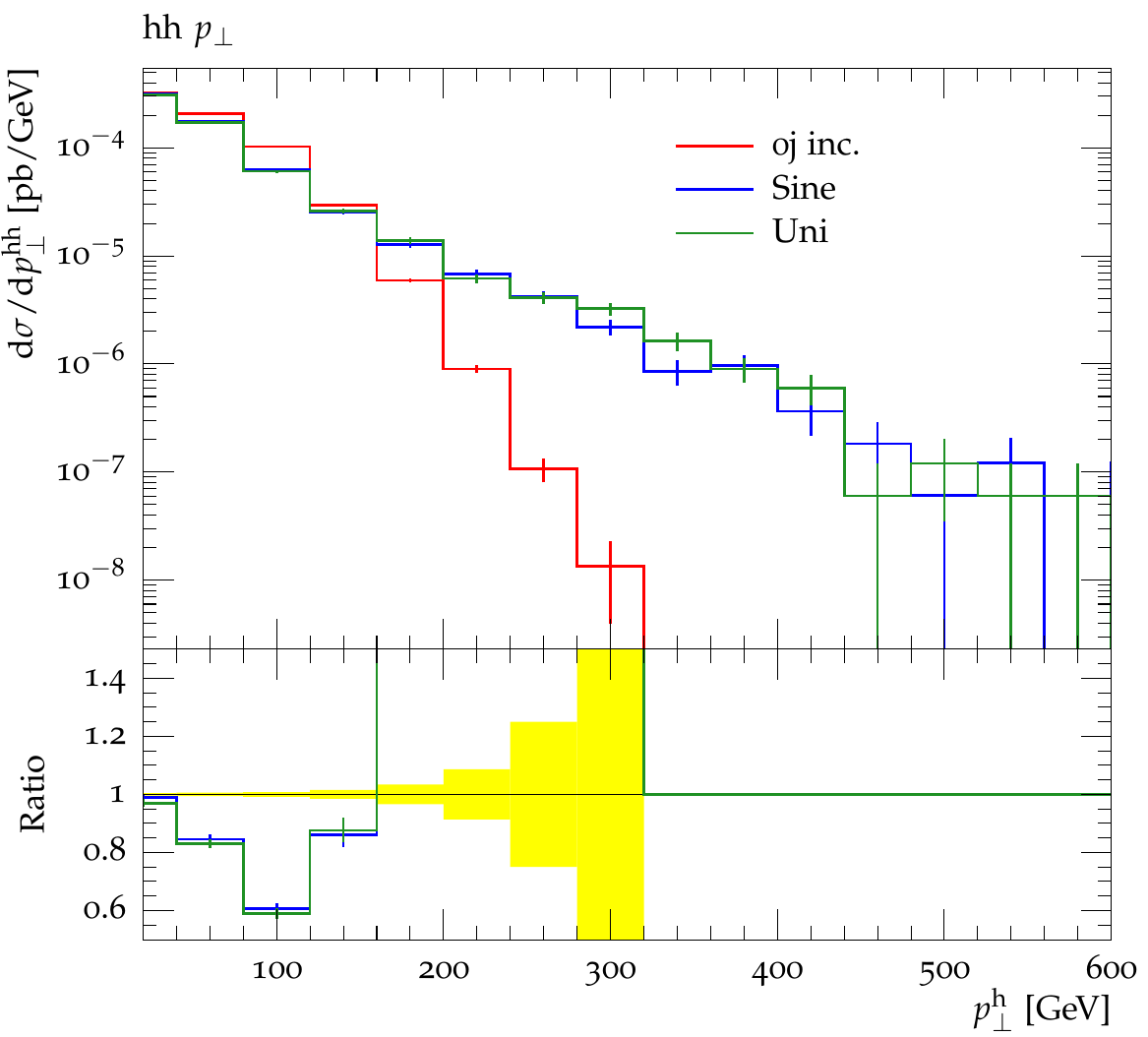}
    \includegraphics[width=0.40\linewidth]{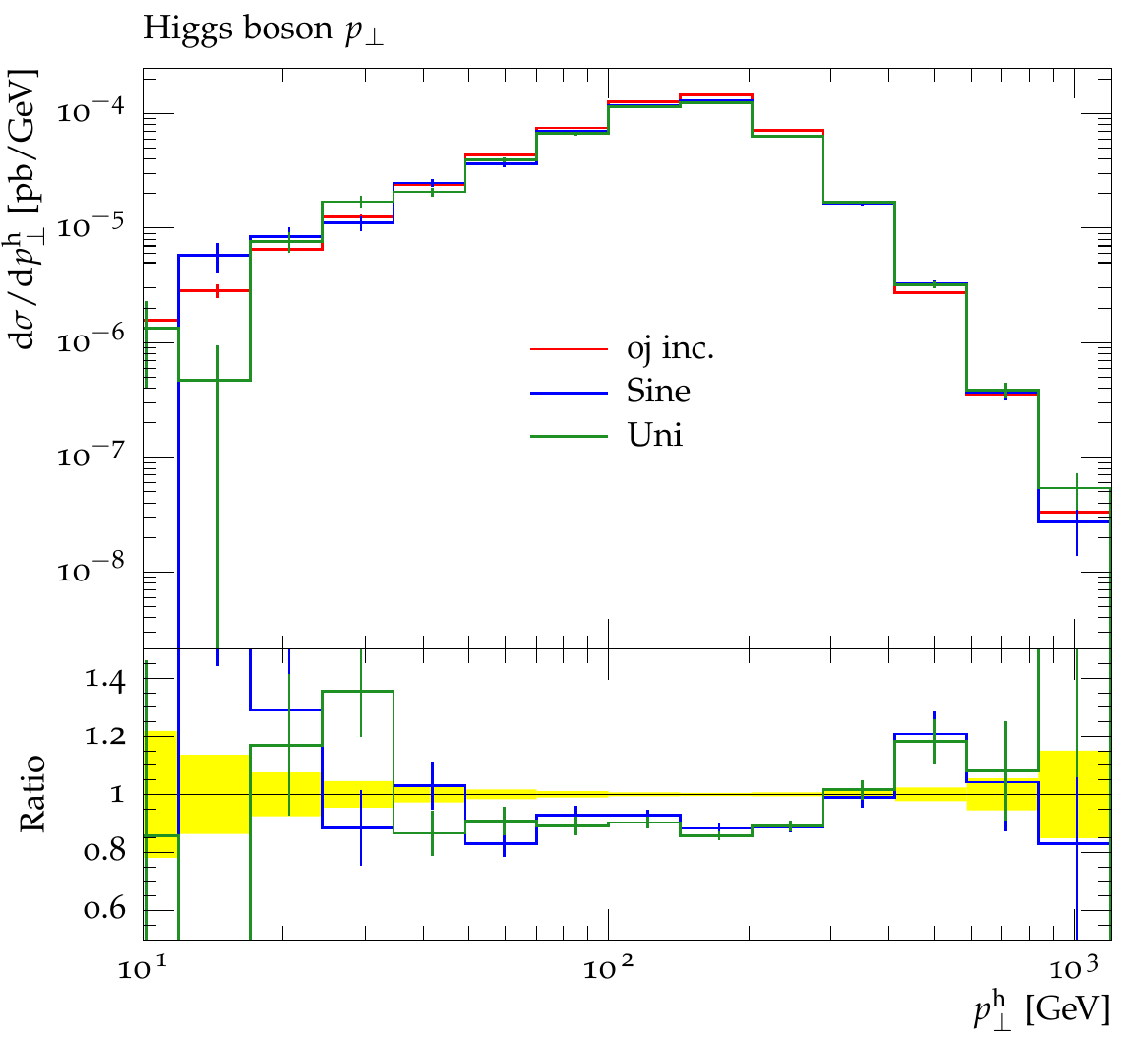}
    \includegraphics[width=0.40\linewidth]{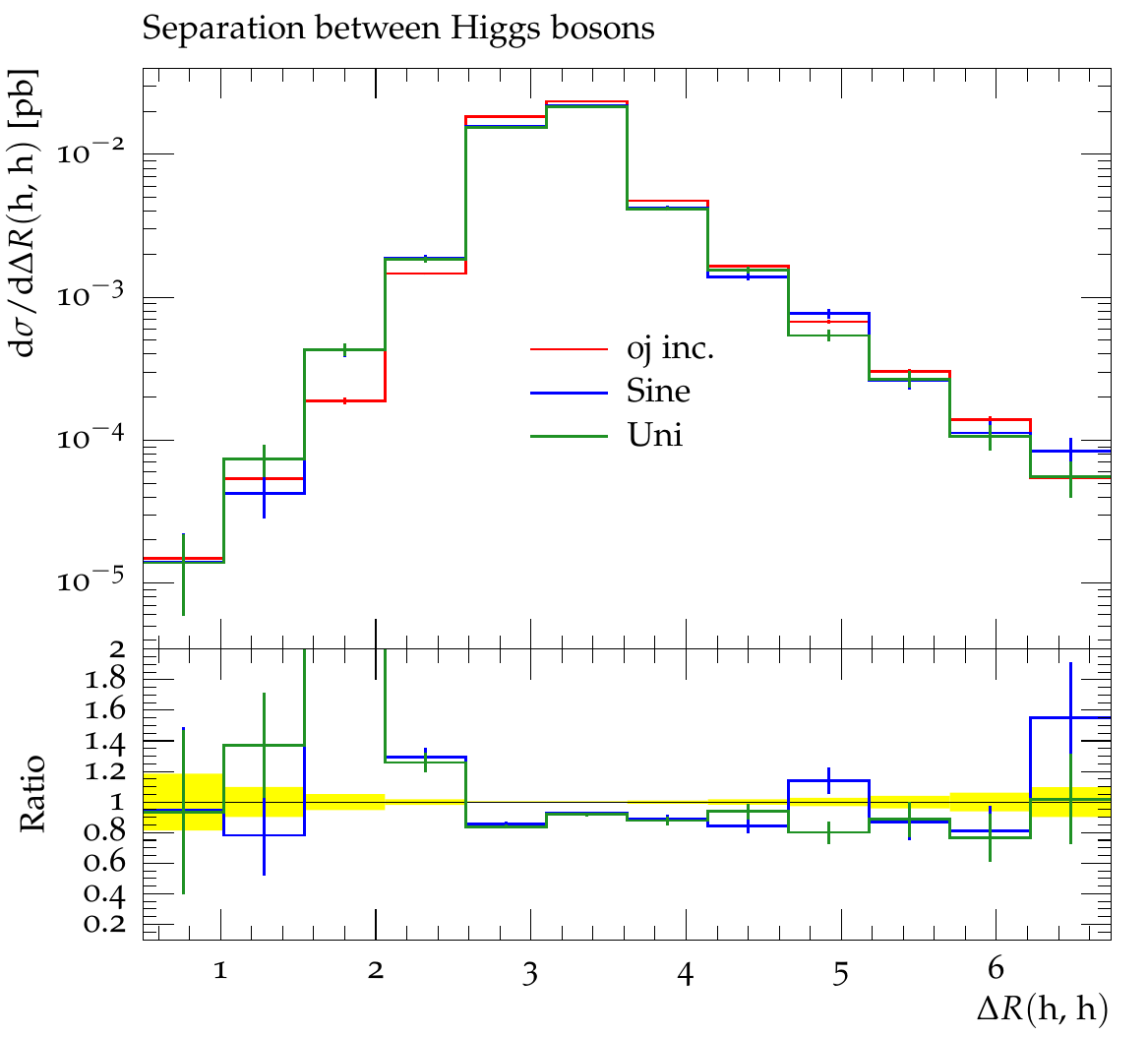}
    \includegraphics[width=0.40\linewidth]{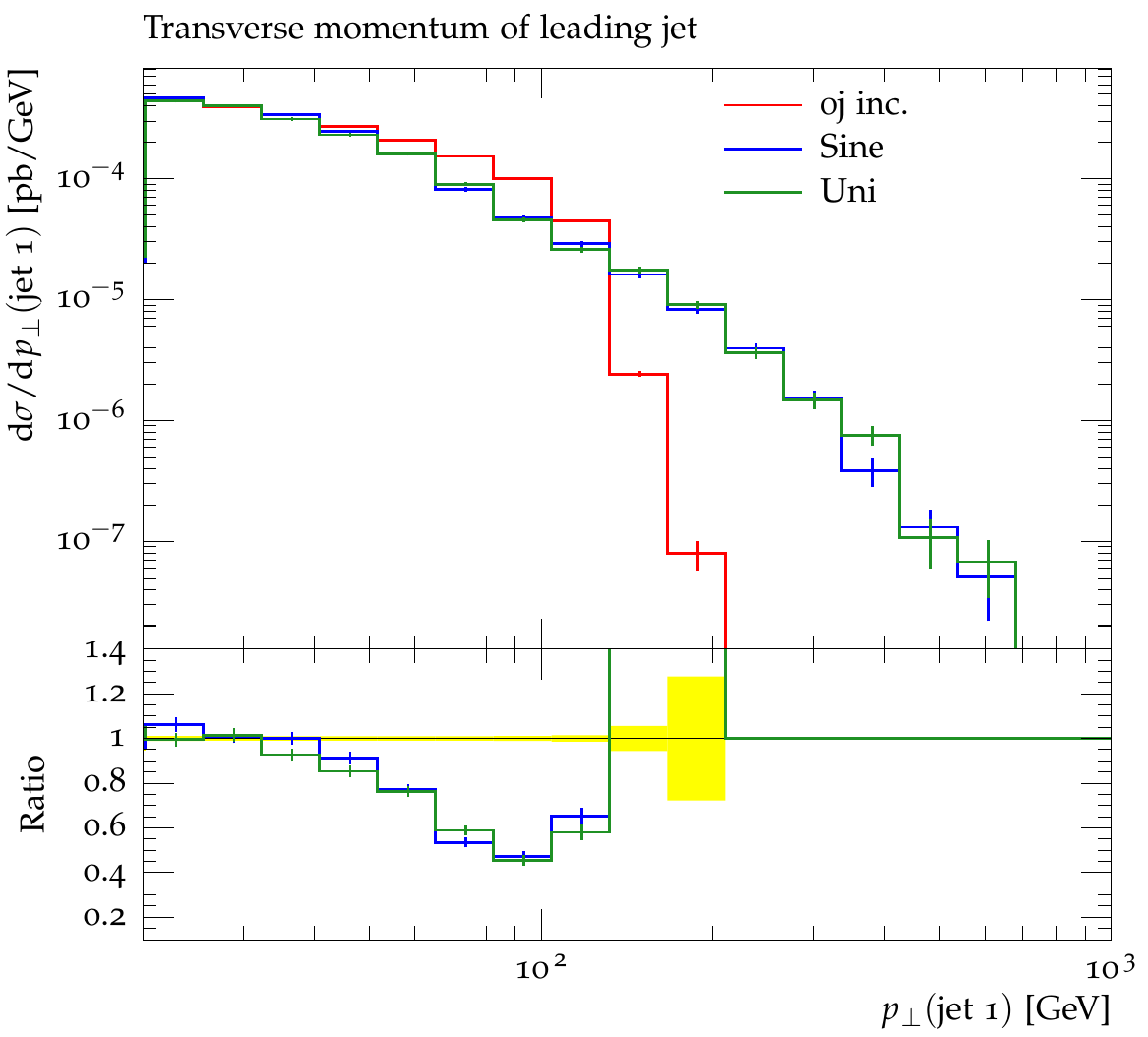}
  \caption{The transverse momentum of
the di-Higgs system and the transverse momentum of a Higgs boson,
$p^{hh}_\perp$ and $p_\perp^h$ respectively (top), the distance between the two Higgs
bosons, $\Delta R(h,h)$, and the $p_\perp$ of the leading jet (bottom). A comparison between the two different smoothing schemes,
`Sinusoidal' and `Uniform' is shown. The clustering parameters were
set to $\bar{E}_{Tclus} =60$~GeV, $\epsilon_{clus}=30$~GeV in both
cases. We also show the un-merged sample (`0j inc.') with $\mu = m_h$, with respect to which the ratio sub-plot is taken.}
  \label{fig:smoothscheme}
\end{figure}

We first examine the effect of the two schemes suggested to
facilitate smooth merging, `Sinusoidal' and `Uniform', for $E_{TClus} = 60$~GeV and $\epsilon_{clus} =
30$~GeV. We compare to the purely showered LO sample (`un-merged')
with $\mu = m_h$. In Fig.~\ref{fig:smoothscheme} we show the transverse momentum of
the di-Higgs system and the transverse momentum of a Higgs boson,
$p^{hh}_\perp$ and $p_\perp^h$ respectively, the distance between the two Higgs
bosons, $\Delta R(h,h)$ and the $p_\perp$ of the leading jet. It is evident by examining the plots that the `Uniform' scheme provides stronger
smoothing than the `Sinusoidal' scheme. Considering that the
disagreement between the un-merged sample and the merged samples in
the merging regions is large, we suggest the use the `Uniform' scheme for merging in the
$hh$ process and we employ this in the rest of this study. 

\begin{figure}
\centering
    \includegraphics[width=0.49\linewidth]{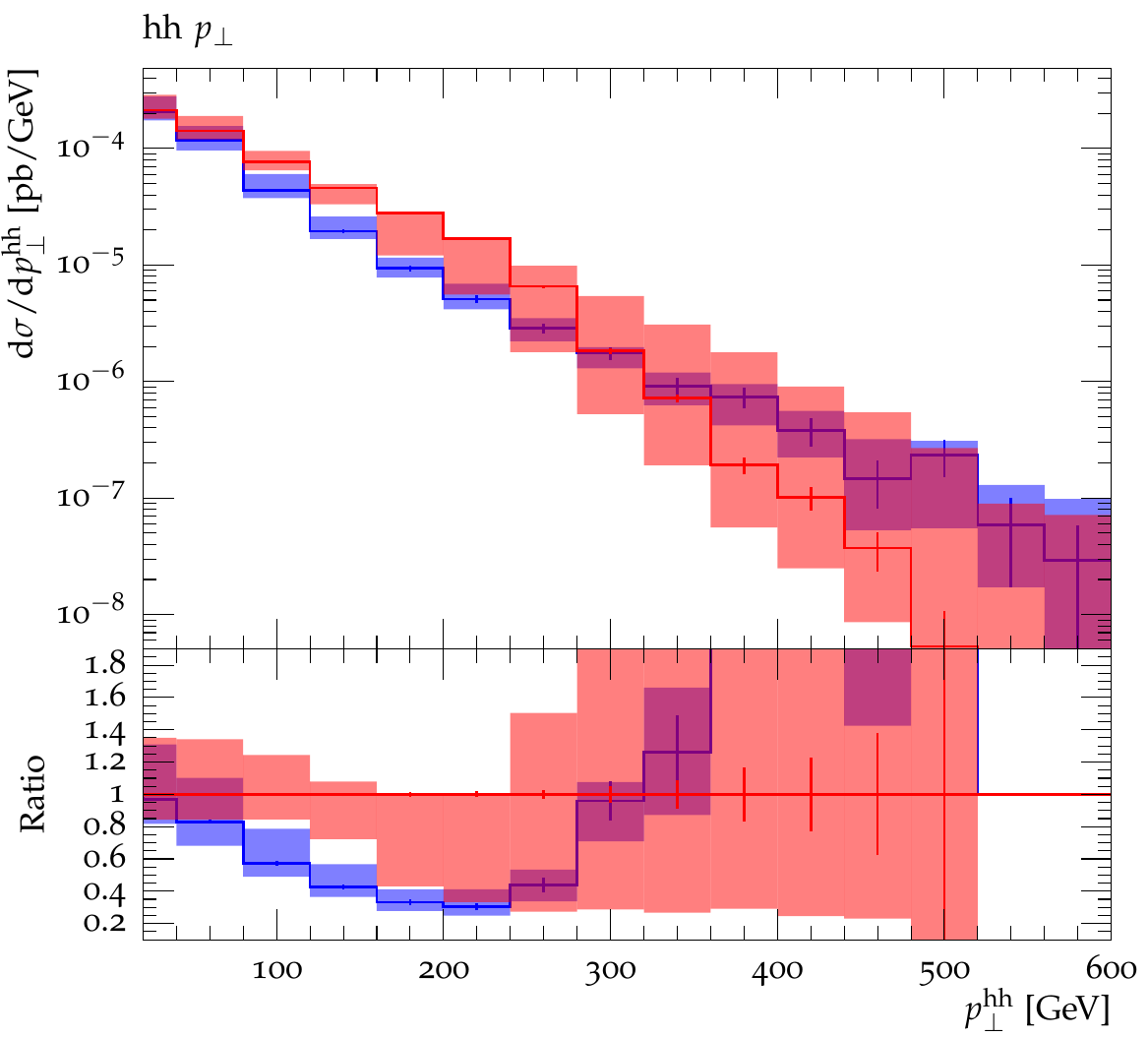}
    \includegraphics[width=0.49\linewidth]{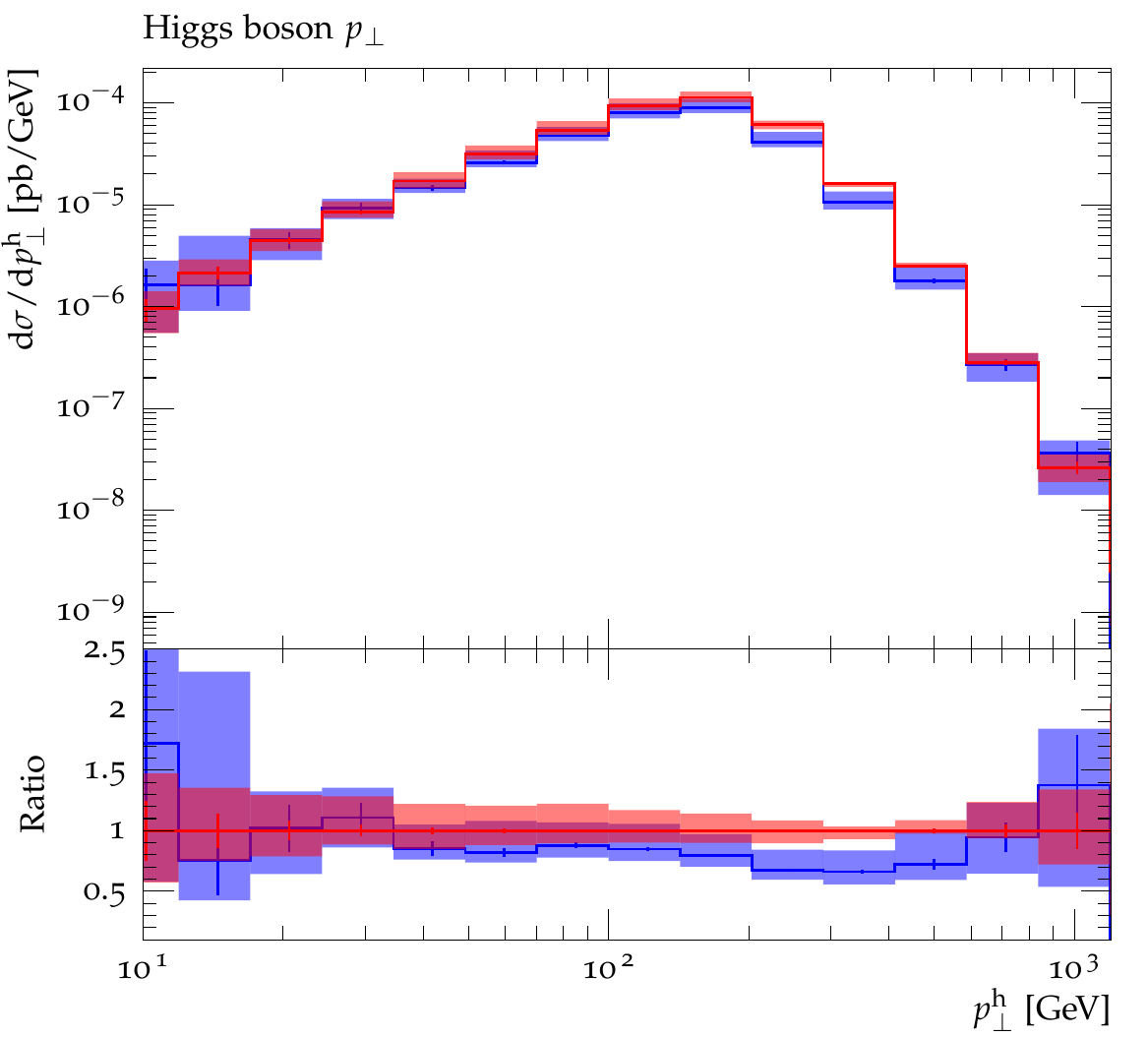}
    \vspace{5mm}
    \includegraphics[width=0.49\linewidth]{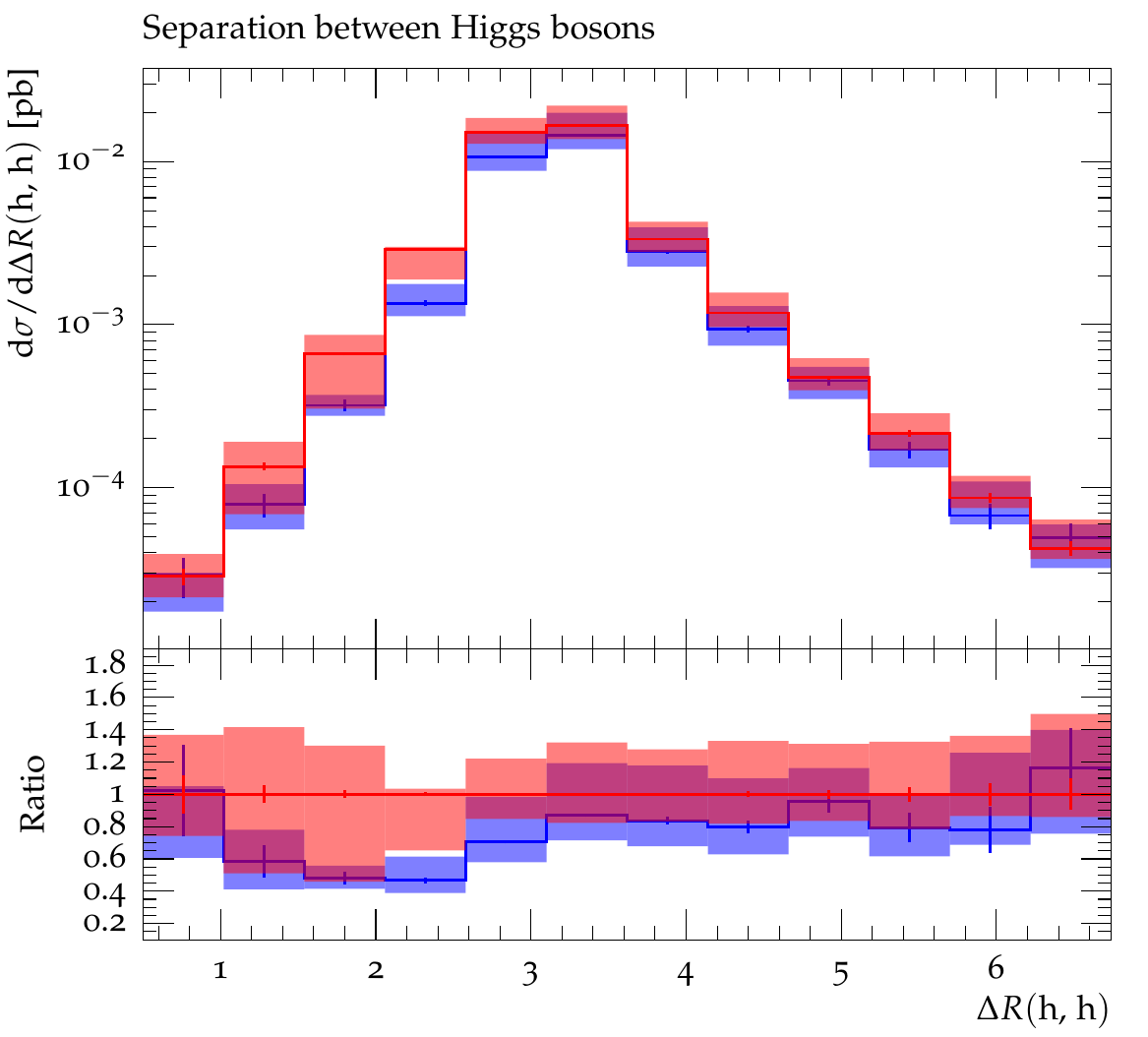}
    \includegraphics[width=0.49\linewidth]{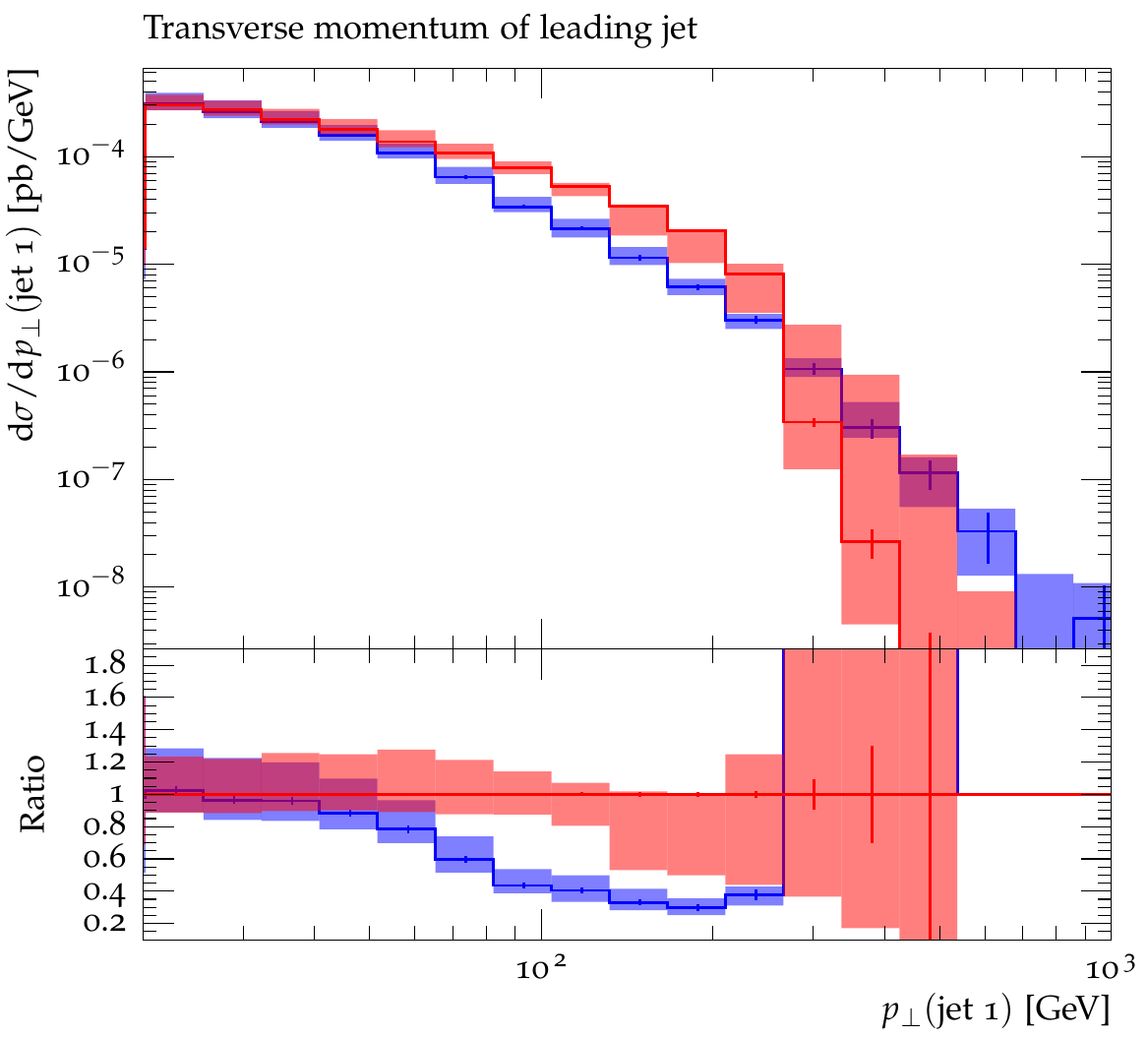}
  \caption{The transverse momentum of
the di-Higgs system and the transverse momentum of a Higgs boson,
$p^{hh}_\perp$ and $p_\perp^h$ respectively (top), the distance between the two Higgs
bosons, $\Delta R(h,h)$, and the $p_\perp$ of the leading jet (bottom). The
merged samples are shown in blue, with the blue line corresponding to
$\mu = 2 ( m_h + p^{hh}_\perp )$ and the un-merged samples
are shown in red, with the red line corresponding to $\mu = 2
m_h$. The bands show the envelope of scale variations between $\mu = m_h +
p^{hh}_\perp$ and $\mu = 4 ( m_h + p^{hh}_\perp)$ for the merged sample and
$\mu = m_h $ and $\mu = 4 m_h$ for the un-merged sample. The merging parameters were chosen to be $\bar{E}_{Tclus} = 60$~GeV, $\epsilon_{clus}=30$~GeV. The ratio sub-plot is taken with
respect to the un-merged sample with $\mu = 2 m_h$.}
  \label{fig:scalevar}
\end{figure}

In Fig.~\ref{fig:scalevar} we examine the effect of the variation of
the scale $\mu$ on various distributions for the merged samples. We
vary the scale $\mu$ between $\mu = m_h +
p^{hh}_\perp$ and $\mu =4 (m_h +p^{hh}_\perp)$, while fixing $\bar{E}_{Tclus}=60$~GeV and
$\epsilon_{clus}=30$~GeV. For comparison, we also show the equivalent un-merged scale variation between $\mu = m_h$ and $\mu =
4m_h$. It is evident that, modulo normalisation differences originating from the
scale variation, the general shapes of the distributions exhibit
reasonable stability over the range of the chosen scales for the
merged sample. More importantly, the scale variation in the
observables $p^{hh}_\perp$, $p_\perp$ of the leading jet and $\Delta R(h,h)$, is substantially reduced with respect to the
leading-order showered samples. This is particularly true in the regions where the
parton shower is not expected to provide a good description of the
additional radiation, namely in the
$\Delta R(h,h) \lesssim \pi$ region and the high-$p_\perp$ regions. These improvements should not
come as a surprise, as the considered observables are leading-order
accurate for the merged sample in those regions, versus leading-logarithmic for the
showered leading-order sample. The distribution of transverse momentum of a single Higgs boson, $p^{h}_\perp$, is not
dominated by the extra radiation and thus the improvement is only marginal. 

\begin{figure}
\centering
    \includegraphics[width=0.49\linewidth]{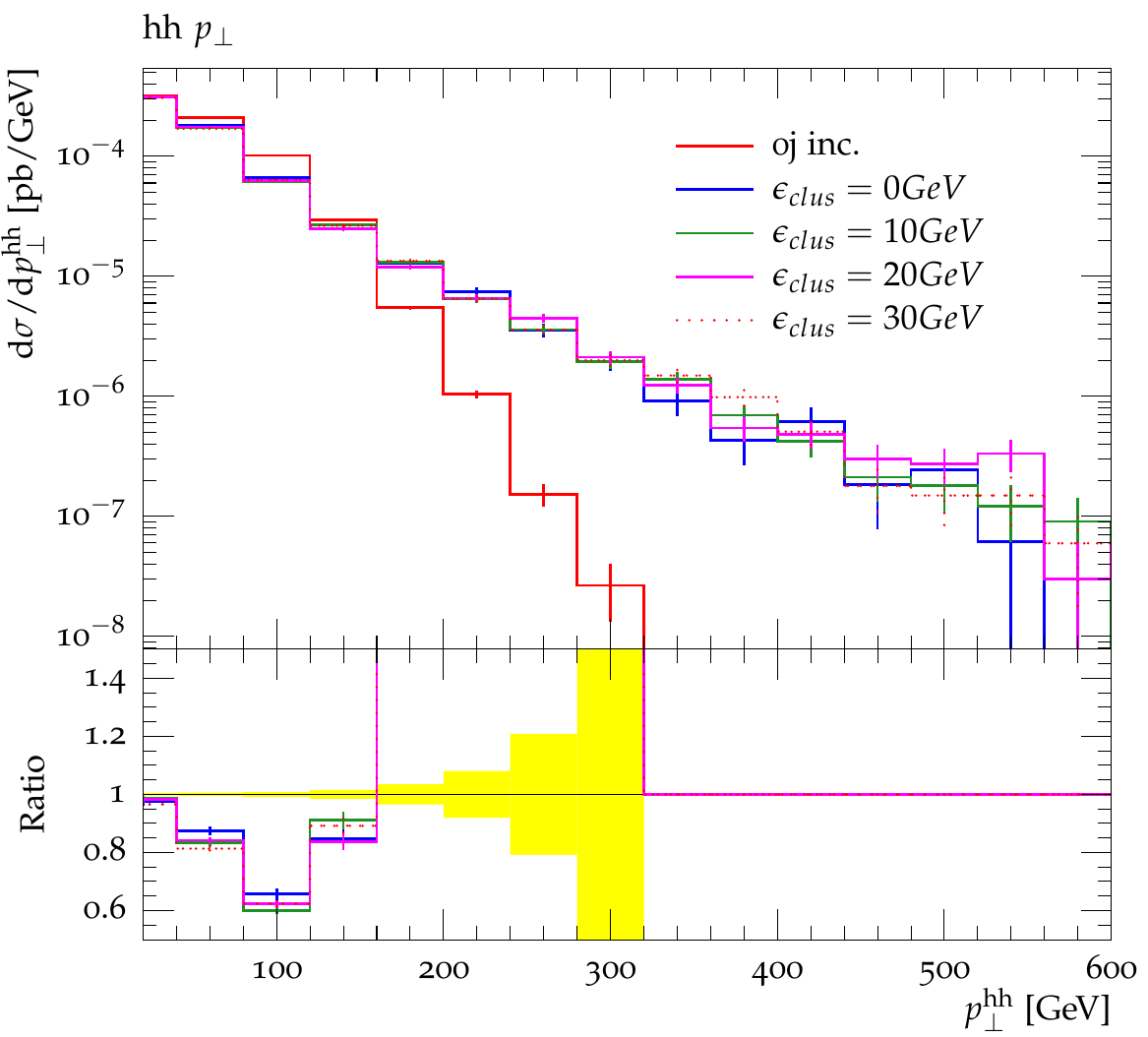}
    \includegraphics[width=0.49\linewidth]{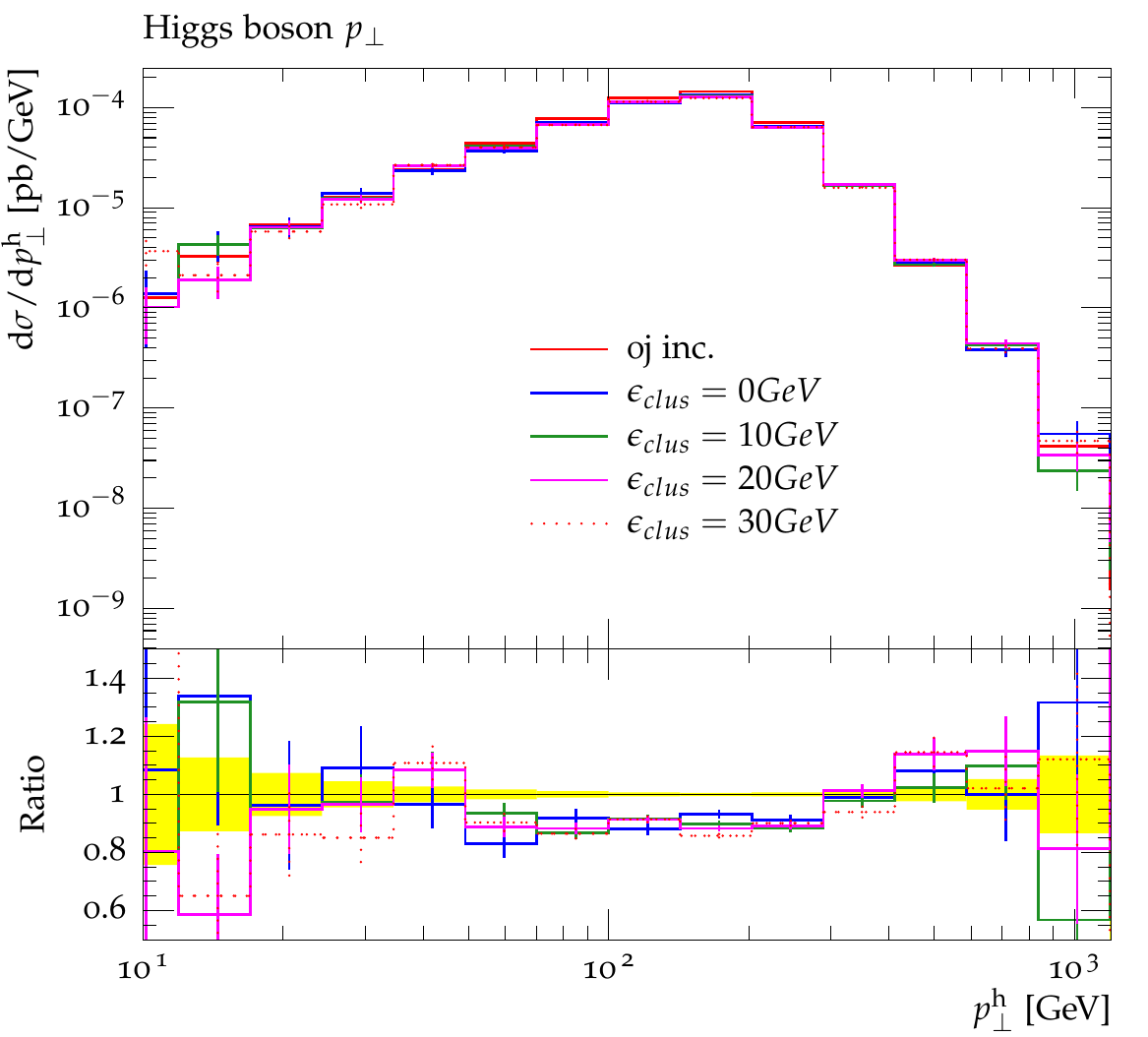}
    \vspace{5mm}
    \includegraphics[width=0.49\linewidth]{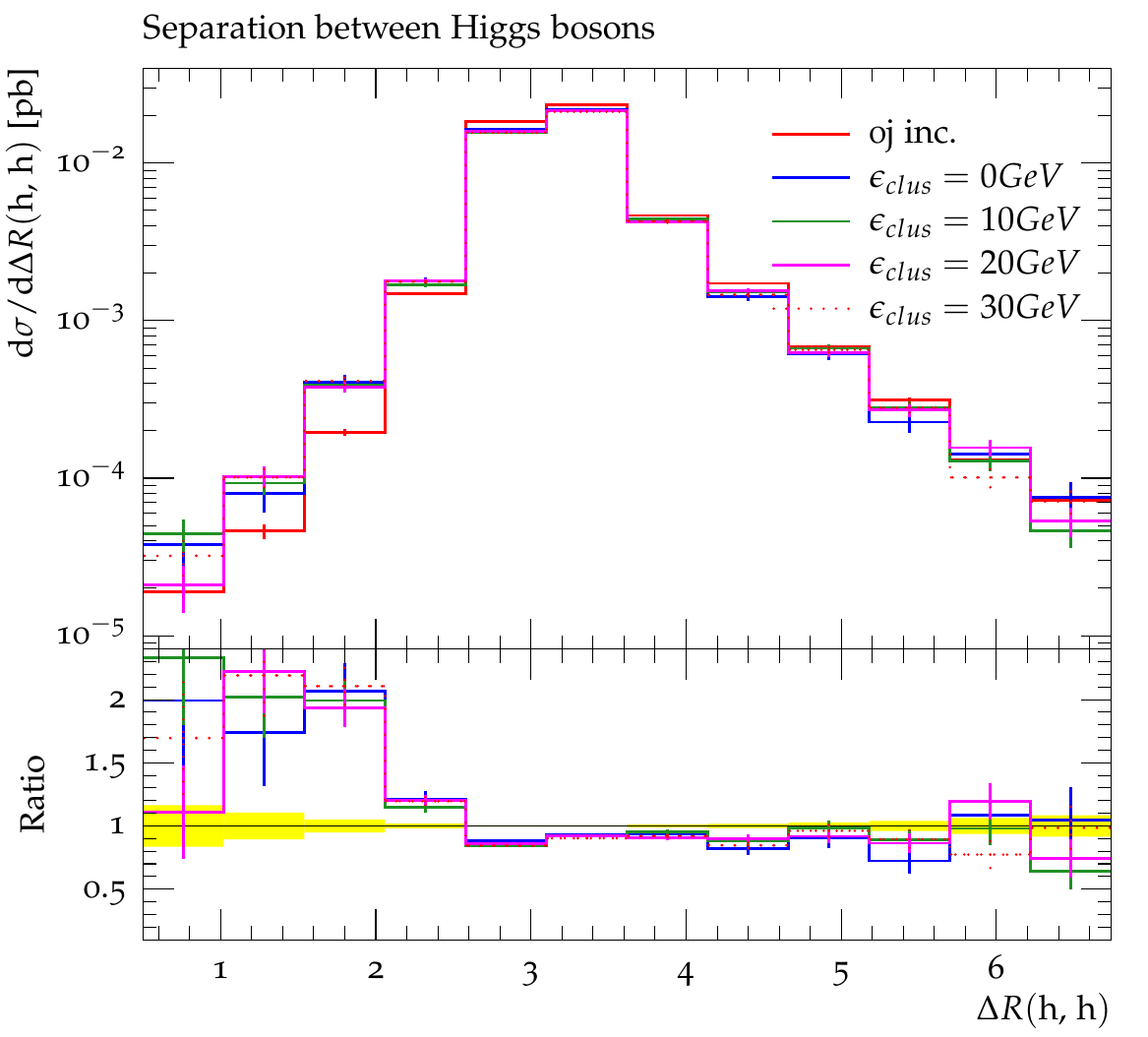}
    \includegraphics[width=0.49\linewidth]{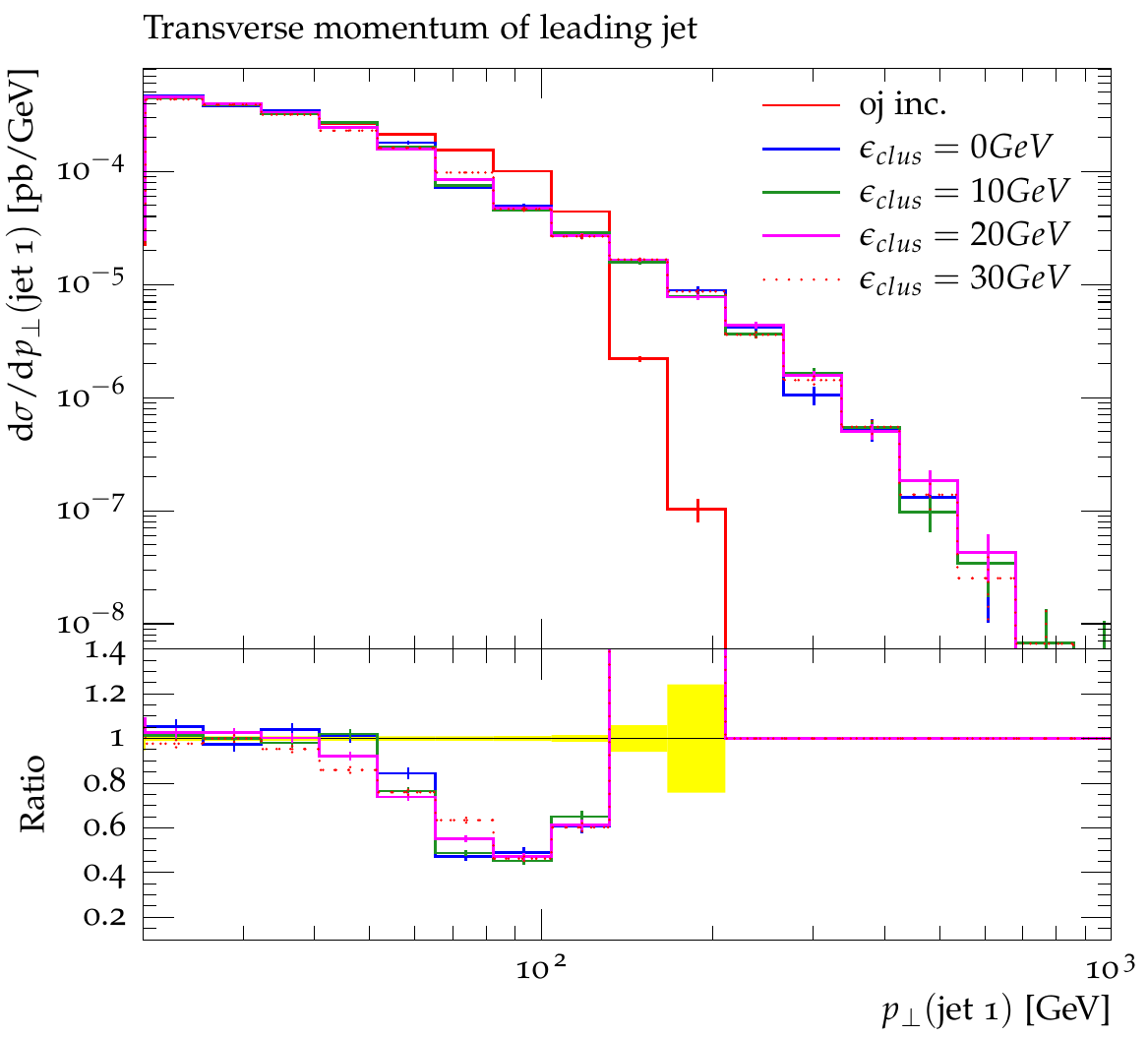}
  \caption{The transverse momentum of
the di-Higgs system and the transverse momentum of a Higgs boson,
$p^{hh}_\perp$ and $p_\perp^h$ respectively (top), the distance between the two Higgs
bosons, $\Delta R(h,h)$, and the $p_\perp$ of the leading jet
(bottom). Different $\epsilon_{clus}$ are chosen and the other
parameters set to $\bar{E}_{Tclus} =
60$~GeV, $\mu=m_h + p_\perp^{hh}$. The ratio sub-plot is taken with
respect to the un-merged sample with $\mu = m_h$ (`0j inc.') and the
yellow bands in the ratio sub-plot represent the Monte Carlo
statistical uncertainty in that sample. }
  \label{fig:epsvar}
\end{figure}

In Fig.~\ref{fig:epsvar} we examine the effect of different
choices of $\epsilon_{clus}$ in the range $[0,30]$~GeV. We again
compare to the un-merged sample with $\mu = m_h$. The average merging scale
was set to $\bar{E}_{Tclus} = 60$~GeV and the scale $\mu$ was set to
$\mu = m_h + p^{hh}_\perp$. Evidently, smoother
merging of the samples can be achieved using higher values of $\epsilon_{clus}$. 

\begin{figure}
\centering
    \includegraphics[width=0.49\linewidth]{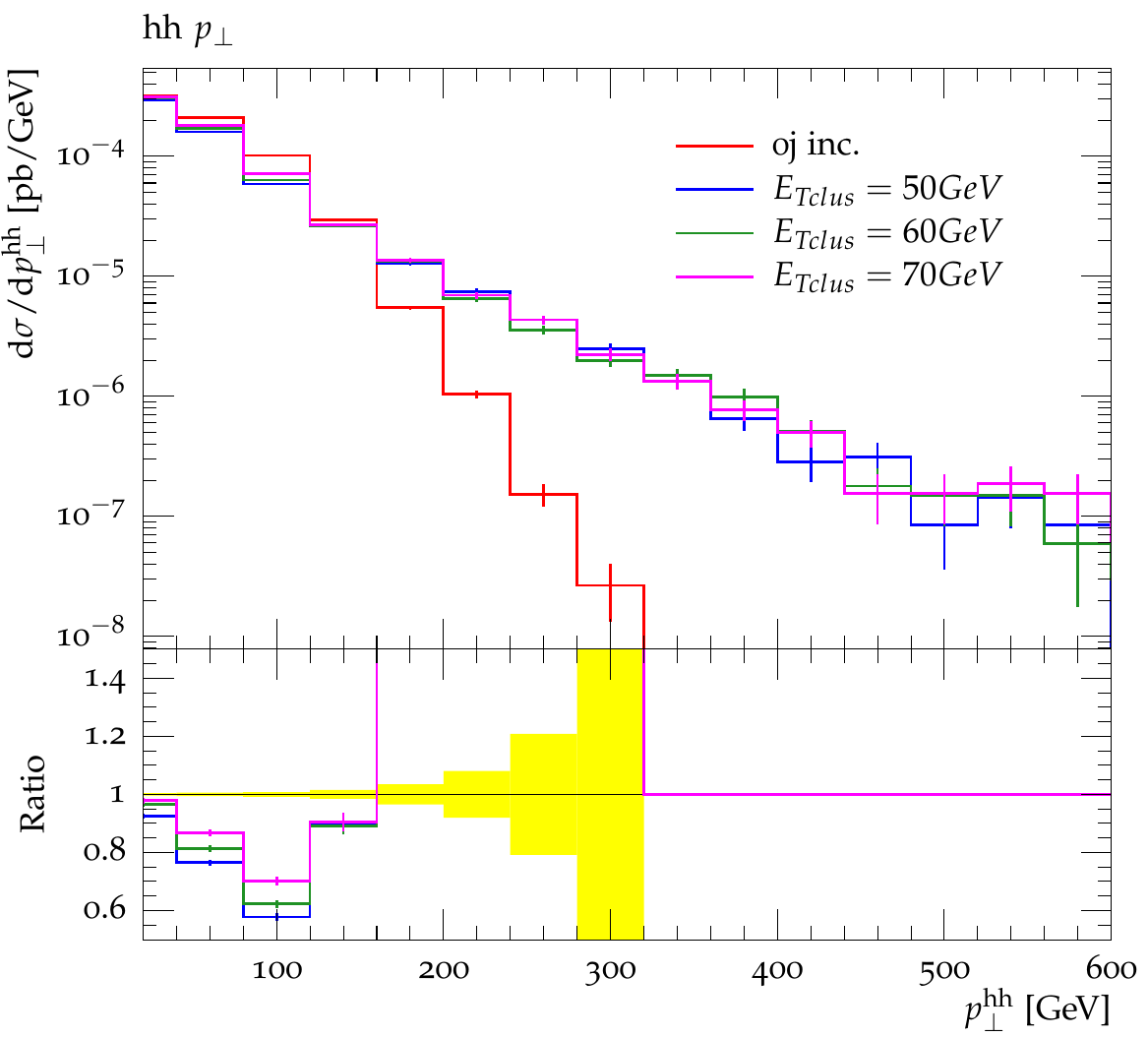}
    \includegraphics[width=0.49\linewidth]{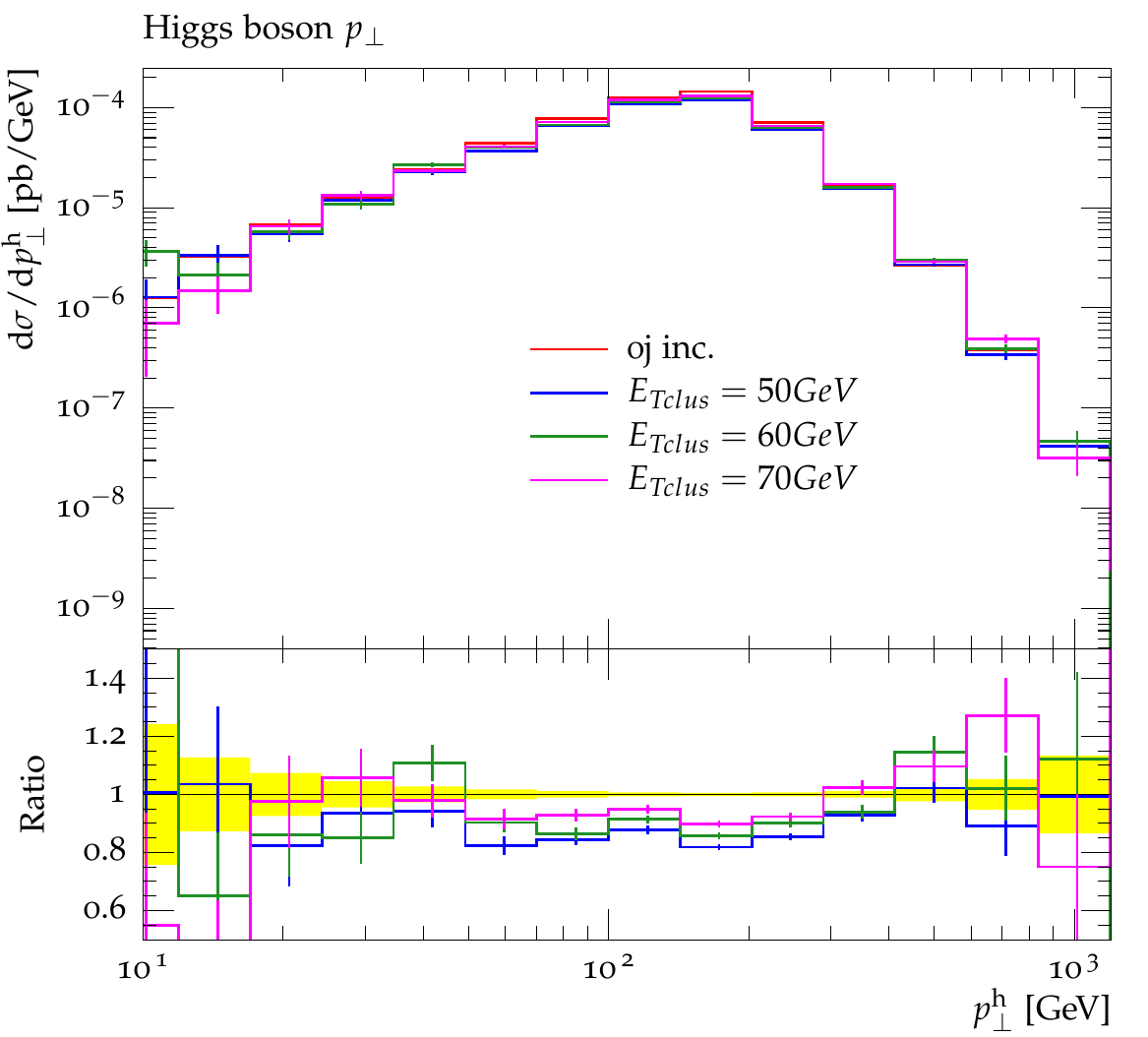}
    \vspace{5mm}
    \includegraphics[width=0.49\linewidth]{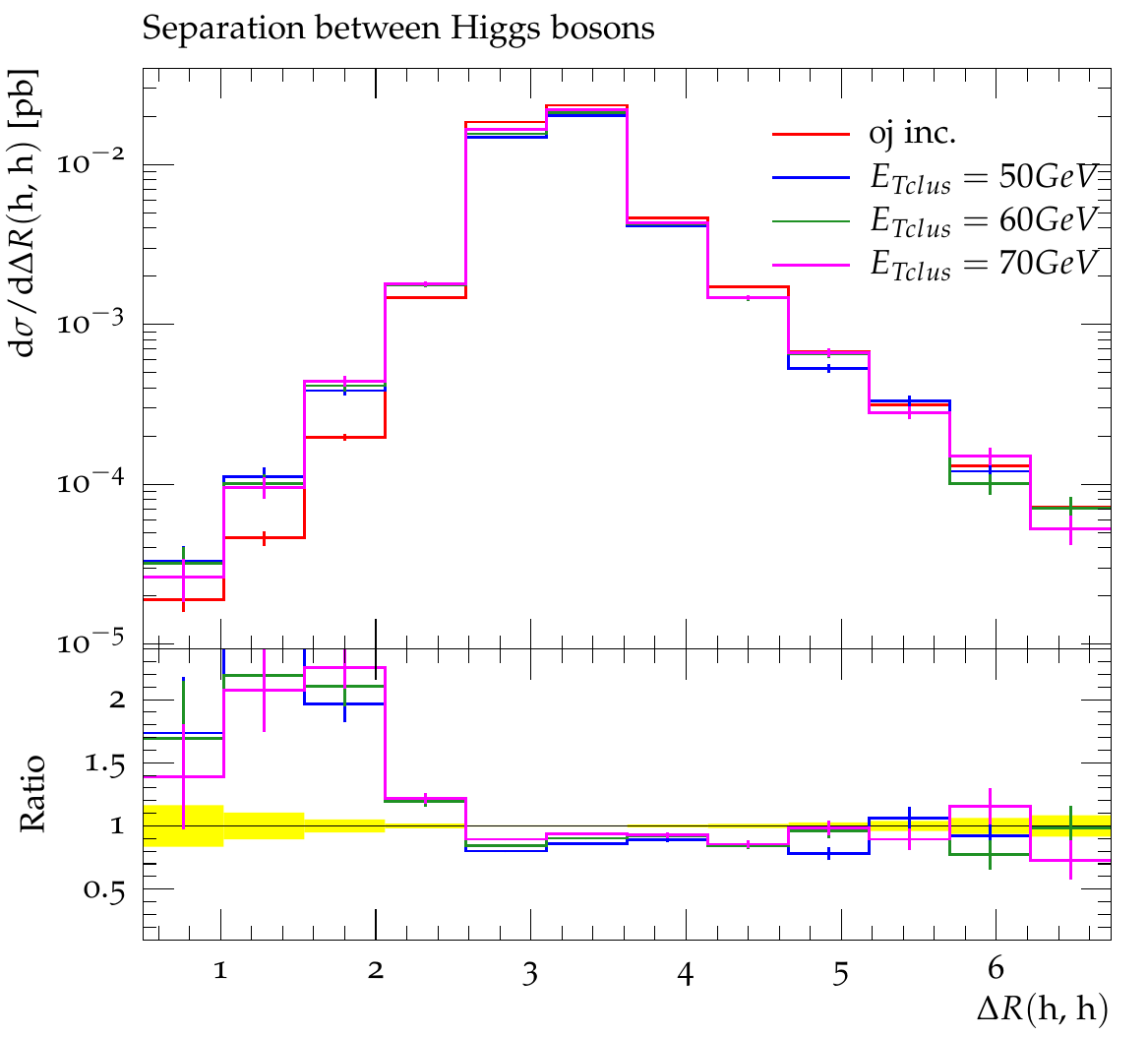}
    \includegraphics[width=0.49\linewidth]{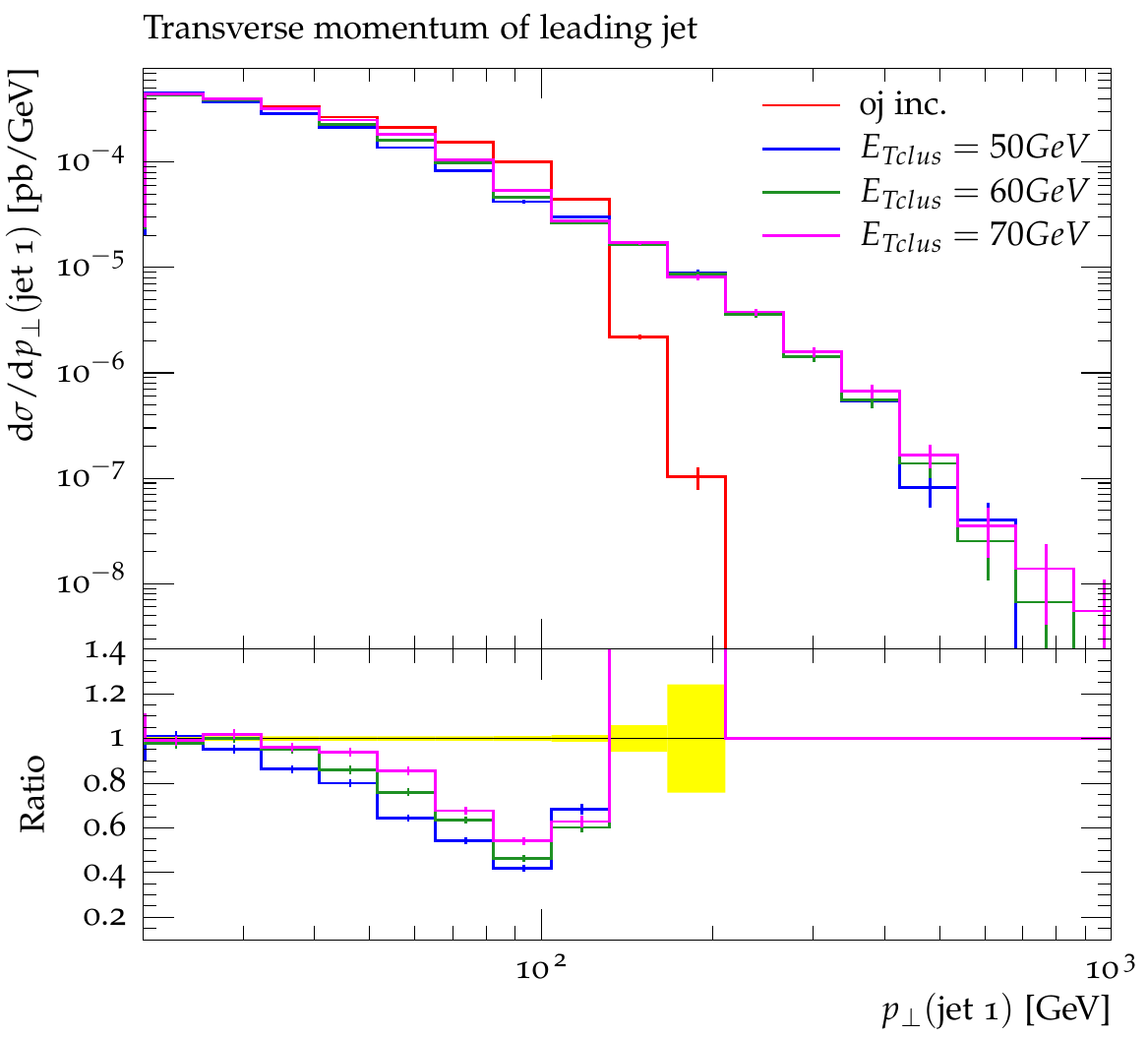}
  \caption{The transverse momentum of
the di-Higgs system and the transverse momentum of a Higgs boson,
$p^{hh}_\perp$ and $p_\perp^h$ respectively (top), the distance between the two Higgs
bosons, $\Delta R(h,h)$, and the $p_\perp$ of the leading jet
(bottom). Different $E_{Tclus}$ are chosen and the other parameters set to $\epsilon_{clus} =
30$~GeV, $\mu=m_h + p_\perp^{hh}$. The ratio sub-plot is taken with
respect to the un-merged sample with $\mu = m_h$ (`0j inc.') and the
yellow bands in the ratio sub-plot represent the Monte Carlo
statistical uncertainty in that sample.}
  \label{fig:etclusvar}
\end{figure}

In Fig.~\ref{fig:etclusvar} we vary the average clustering
scale, $\bar{E}_{Tclus}$, while keeping $\mu = m_h + p^{hh}_\perp$ and
$\epsilon_{clus} = 30$~GeV. We again compare to the un-merged sample
with $\mu = m_h$. Evidently, a relatively large systematic
uncertainty comes from varying $\bar{E}_{Tclus}$. This is due to the
fact that changing $\bar{E}_{Tclus}$ alters the regions that the parton shower and matrix element calculation
contribute in. For a lower $\bar{E}_{Tclus}$, the transition
is smoother and we will use $\bar{E}_{Tclus} = 50$~GeV for the
phenomenological studies of the next section. We will contrast this to
$\bar{E}_{Tclus}=70$~GeV where appropriate.  

\begin{figure}
\centering
    \includegraphics[width=0.49\linewidth]{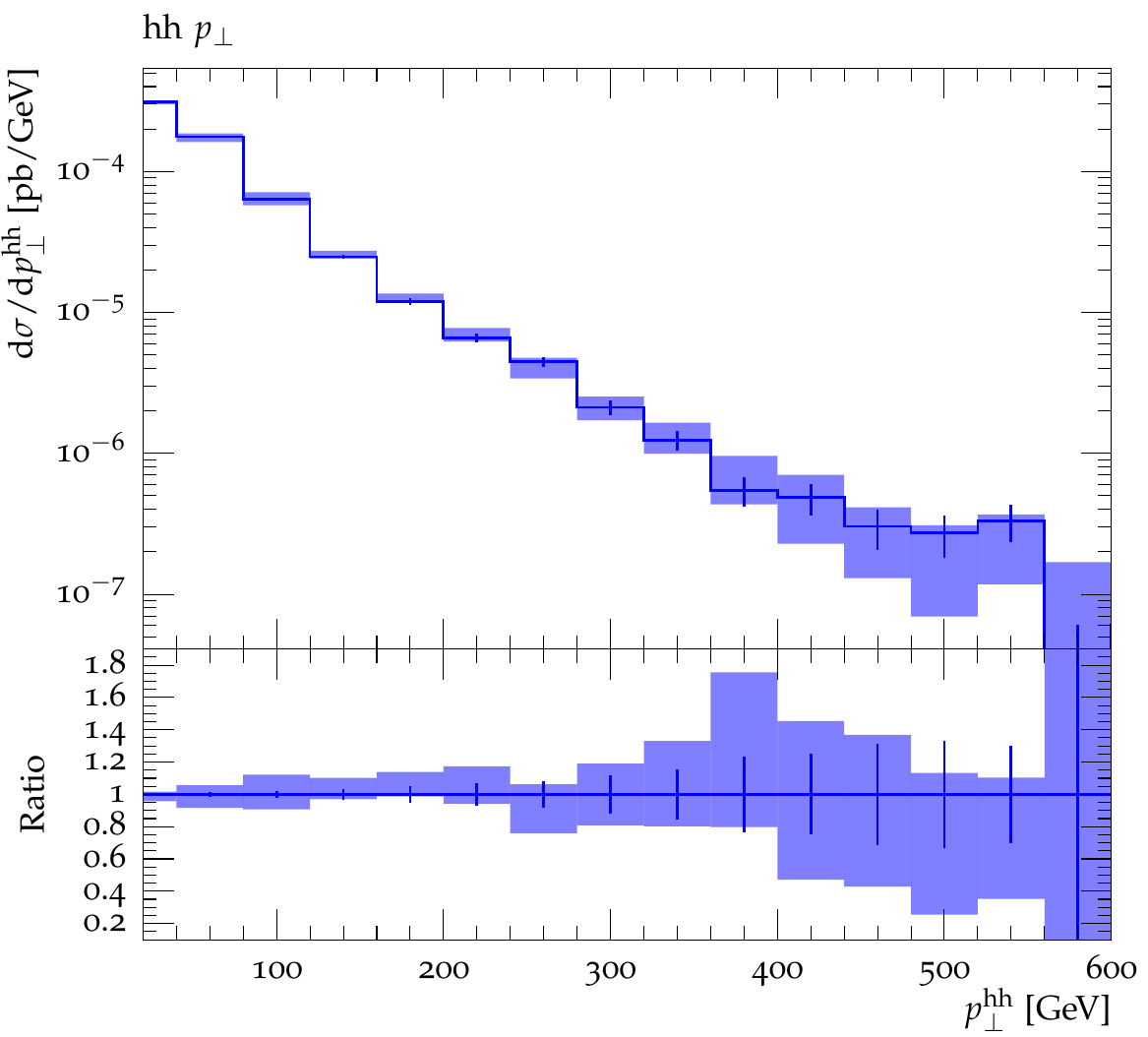}
    \includegraphics[width=0.49\linewidth]{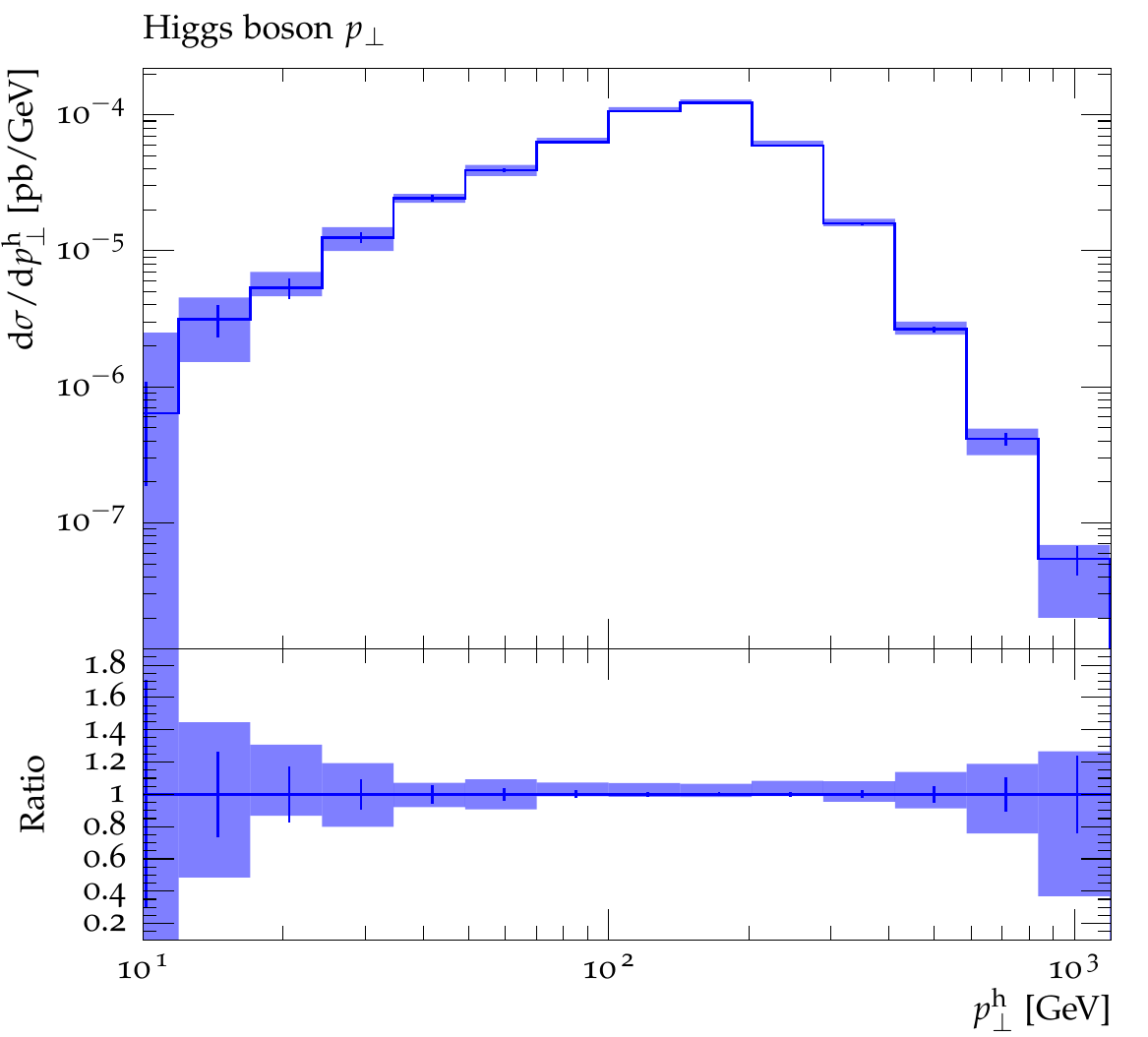}
    \vspace{5mm}
    \includegraphics[width=0.49\linewidth]{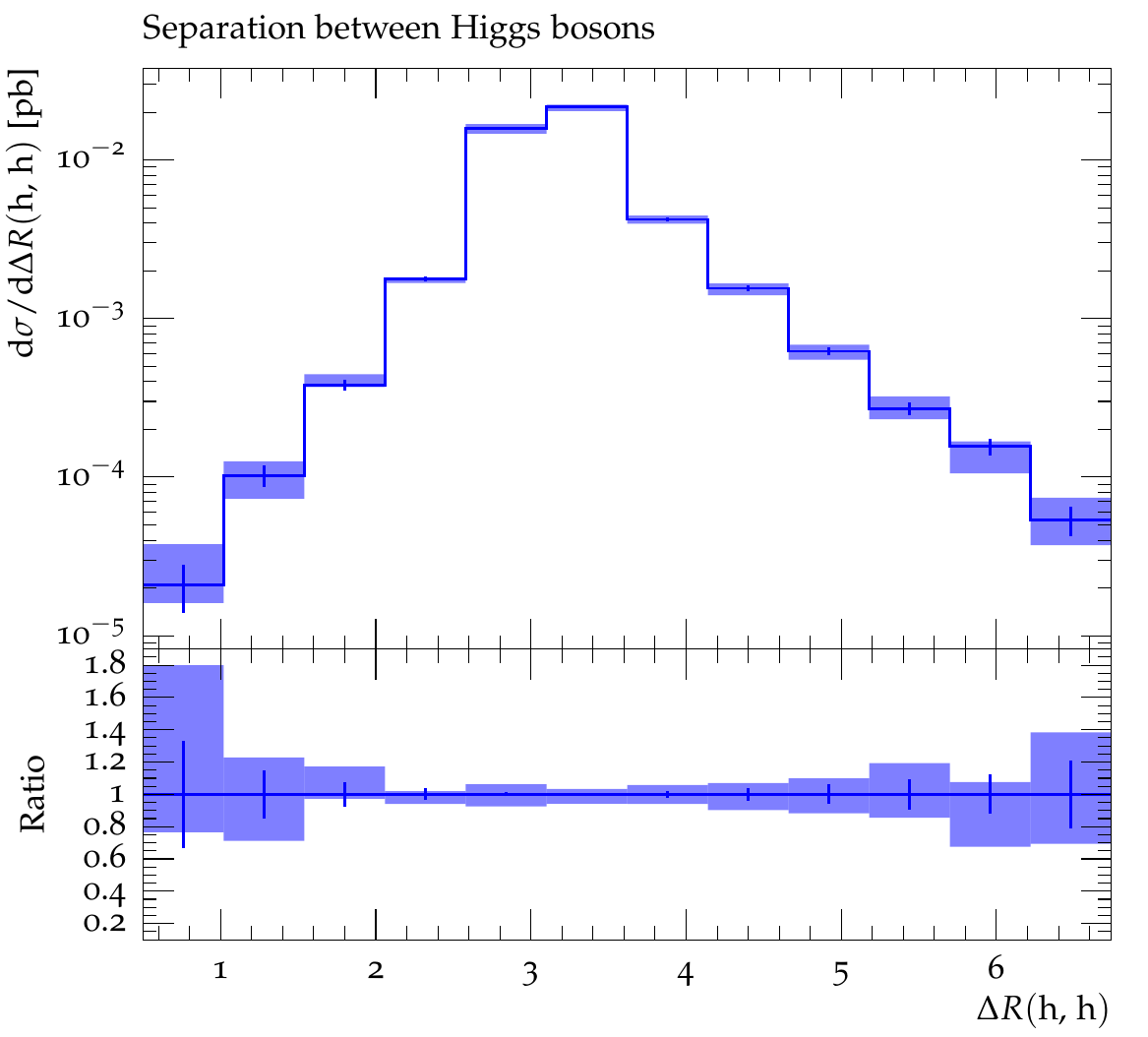}
    \includegraphics[width=0.49\linewidth]{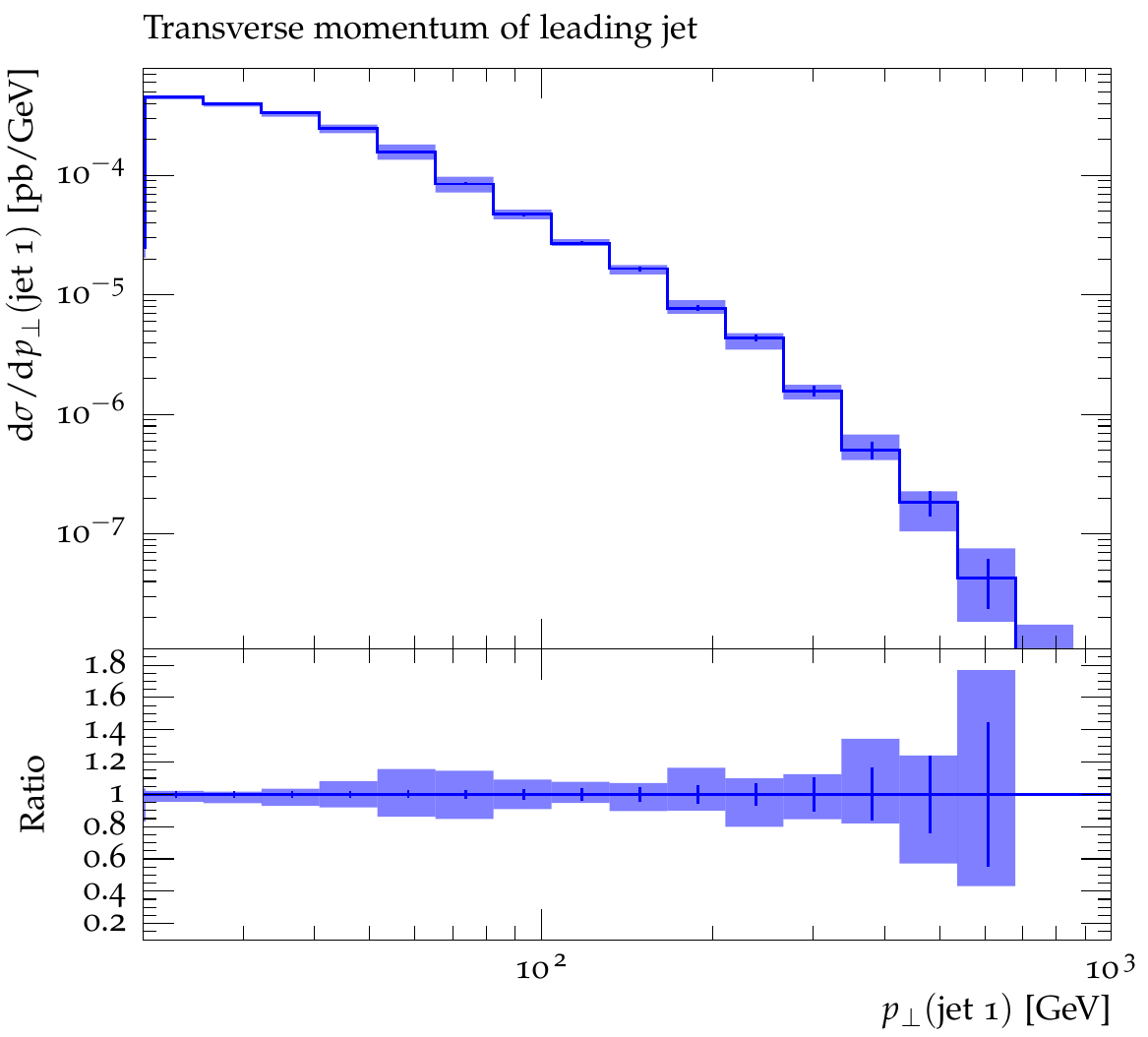}
  \caption{The transverse momentum of
the di-Higgs system and the transverse momentum of a Higgs boson,
$p^{hh}_\perp$ and $p_\perp^h$ respectively (top), the distance between the two Higgs
bosons, $\Delta R(h,h)$ and the $p_\perp$ of the leading jet (bottom). The uncertainty envelope is constructed for
$E_{Tclus} \in [50, 70]$~GeV and $\epsilon_{clus} \in [10,30]$~GeV,
$\mu=m_h + p_\perp^{hh}$. The ratio sub-plot is taken with
respect to the case where $E_{Tclus} = 60$~GeV and $\epsilon_{clus} =
20$~GeV. The error bars represent the Monte Carlo statistical uncertainty for that set
of parameters. }
  \label{fig:envelope}
\end{figure}

Finally, we show the envelope of the variation of both the merging scale
$\bar{E}_{Tclus}$ (in $[50, 70]$~GeV) and the variation scale
$\epsilon_{clus}$ (in $[10, 30]$~GeV) in Fig.~\ref{fig:envelope}. In
all of the samples, the scale was set to $\mu = m_h + p^{hh}_\perp$. The ratio sub-plot in the Figure is taken with respect to the case where
$E_{Tclus} = 60$~GeV and $\epsilon_{clus} = 20$~GeV and the error bars
represent the Monte Carlo statistical uncertainty on that
sample. For the $p^{hh}_\perp$ distribution, the uncertainty due to the variation of
$E_{Tclus}$ and $\epsilon_{clus}$ is $\mathcal{O}(20\%)$ up to
$p^{hh}_\perp \sim 250$~GeV and grows to over $\sim 40\%$ at higher
values. The $p^h_\perp$ distribution exhibits variations of
$\mathcal{O}(10\%)$ or less up to $\sim 400$~GeV and down to $\sim
30$~GeV. For the $\Delta R(h,h)$, the uncertainty is
$\mathcal{O}(20\%)$ in $\Delta R(h,h) \in [1.5,5]$ and close to 40\%
for $\Delta R(h,h) \sim 1$. Conclusions cannot
be made for values outside this range since the samples are
constrained by statistical fluctuations. The $p_\perp$ distribution of
the hardest jet shows below $\mathcal{O}(10\%)$ variations at low $p_\perp$,
which grow to $\mathcal{O}(30\%)$ around the region where the merging
scale becomes significant, $[50, 70]$~GeV, and are then reduced to $\mathcal{O}(10-20\%)$
variations up to $p_\perp \sim 300$~GeV.  

\section{Phenomenological implications}\label{sec:pheno}

It is important to examine the implications of the merging on
realistic phenomenological
analyses of Higgs boson pair production at the LHC. We do this by focussing on an example of a decay channel with a
relatively large branching ratio, $hh \rightarrow (b\bar{b})
(\tau^+ \tau^-)$. This has been examined in detail
in~\cite{Dolan:2012rv, Baglio:2012np, Barr:2013tda}. We do
not attempt here to perform a detailed signal versus background
study; instead, we wish to show the magnitude of the effect of using
the merged sample in a realistic analysis. We only focus on the
top-anti-top background which will constitute the largest component of the
irreducible background, via
\begin{equation}
pp \rightarrow t \bar{t} \rightarrow (\tau^- \bar{\nu}_\tau b) (
\tau^+ \nu_\tau \bar{b}) \;.
\end{equation}
We consider the case of a 14~TeV LHC, and normalise all $hh$
inclusive cross sections to the NNLO cross section obtained within
the effective theory in~\cite{deFlorian:2013jea}, $\sigma^{NNLO}_{hh}= 40.2$~fb. We consider
four different samples, un-merged with scales set to $\mu = m_h$
and $\mu = 2m_h$ and merged with scales $\mu = m_h +
p^{hh}_\perp$ and $\mu = 2(m_h+p^{hh}_\perp)$. The merging parameters were
fixed to $E_{Tclus}=50$~GeV (or $70$~GeV) and $\epsilon_{clus} = 30$~GeV. The $t\bar{t}$ background was
generated via \texttt{aMC@NLO}~\cite{Frixione:2010ra, Frederix:2011zi}
along with the decays, and was assumed to have a total cross section
of $\sigma_{t\bar{t}} = 900$~pb~\cite{Ahrens:2011px, Czakon:2013goa}. Showering and hadronization were performed
using \texttt{HERWIG++}, and the simulation of the underlying event was
included via multiple secondary parton interactions~\cite{Bahr:2008dy}. We follow the basic analysis steps as
given in~\cite{Dolan:2012rv}: we assume 80\% $\tau$-reconstruction efficiency
with negligible fake rate\footnote{Thus, we do not consider any mistagging
backgrounds, which could be potentially important.} and require two
$\tau$-tagged jets with at least $p_\perp > 20$~GeV. We require that the taus, taken from the Monte Carlo truth, reproduce the Higgs mass within a 50~GeV
window, to account for the reconstruction smearing, as done
in~\cite{Dolan:2012rv}. We use the Cambridge-Aachen jet algorithm available in the
\texttt{FastJet} package~\cite{Cacciari:2011ma, Cacciari:2005hq} with a radius
parameter $R=1.4$ to search for so-called `fat jets'. We require the
existence of one fat jet in the event satisfying the mass-drop
criteria as done in the $hV$ study in Ref.~\cite{Butterworth:2008iy}. We require the two hardest `filtered' sub-jets to
be b-tagged\footnote{Bottom-jet tagging was performed by setting the
bottom mesons to stable in the \texttt{HERWIG++} event generator.} and to be central ($|\eta| < 2.5$) and the filtered fat jet
to be in $(m_h -25~\mathrm{GeV}, m_h + 25~\mathrm{GeV})$. The b-tagging efficiency
was taken to be 70\%, again with negligible fake rate for the sake of
simplicity. We require a loose cut on the
transverse momentum of the fat jet (after filtering) that satisfies the above criteria,
$p_\perp^{\mathrm{fat}} > 100$~GeV. This is done to maintain a sufficient number of events to
examine the change of efficiencies with respect to other cuts. We also
apply a transverse momentum cut on the $\tau^+\tau^-$ system of equal
magnitude, $p_\perp^{\tau\tau} > 100$~GeV. 

\begin{figure}
\centering
    \includegraphics[width=0.44\linewidth]{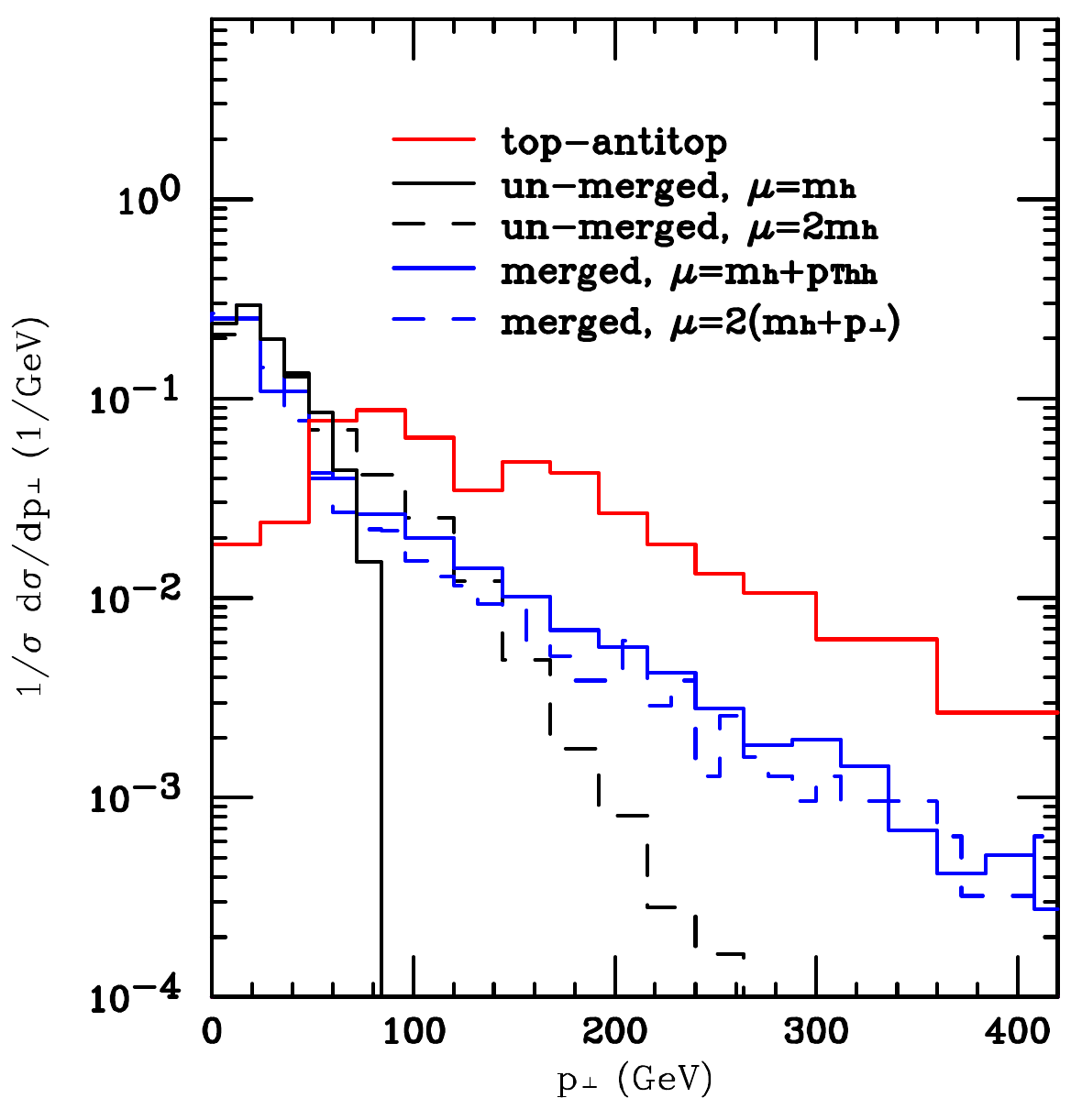}
    \includegraphics[width=0.45\linewidth]{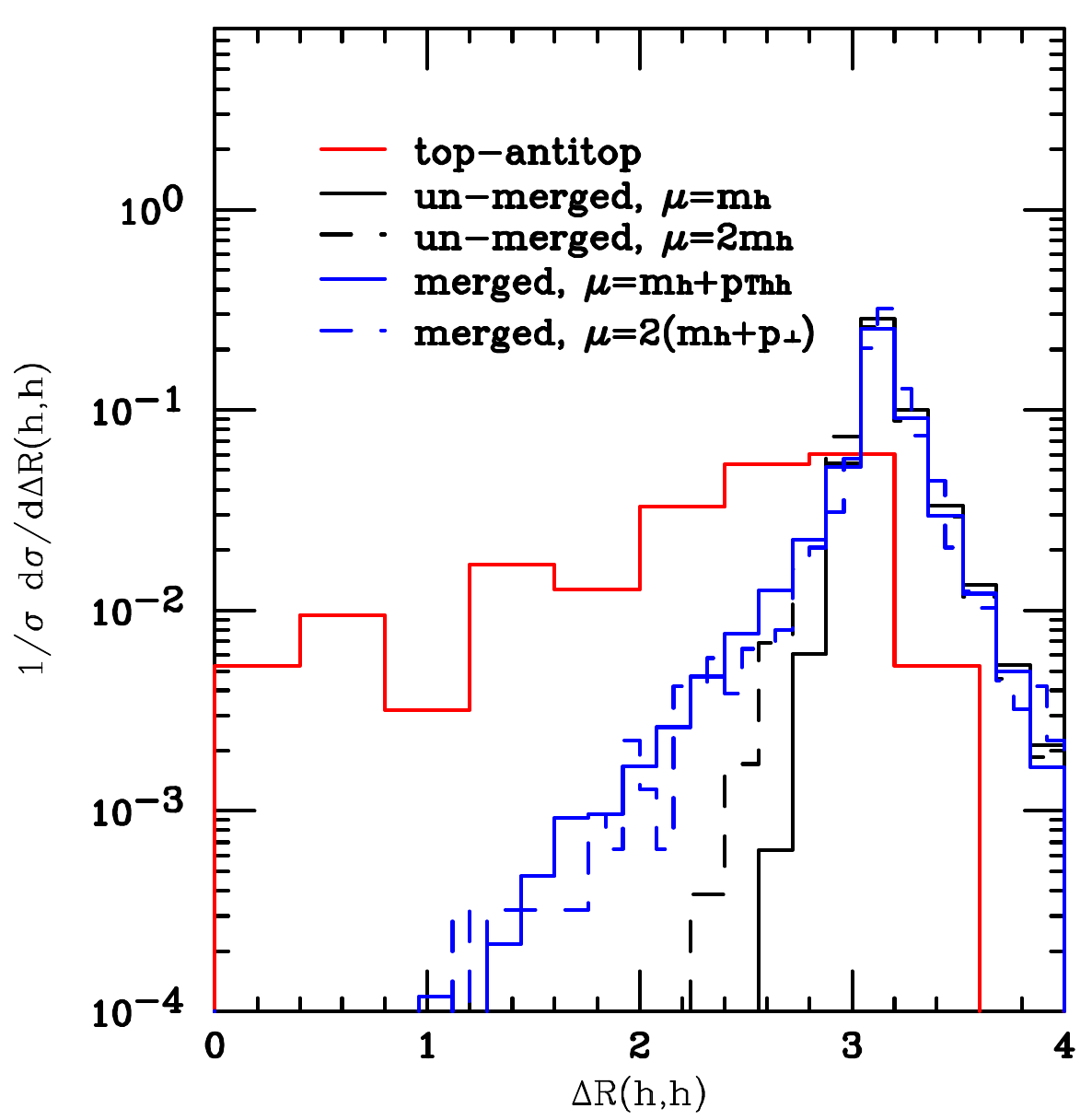}
  \caption{The reconstructed transverse momentum of the Higgs boson
    pair (left) and the distance between the reconstructed Higgs bosons
    (right) resulting from the analysis outlined in the main text for
    the different signal samples (merged or un-merged) and the
    top-anti-top background.}
  \label{fig:observables}
\end{figure}

\begin{figure}
\centering
\hspace{-3.0mm}
    \includegraphics[width=0.45\linewidth]{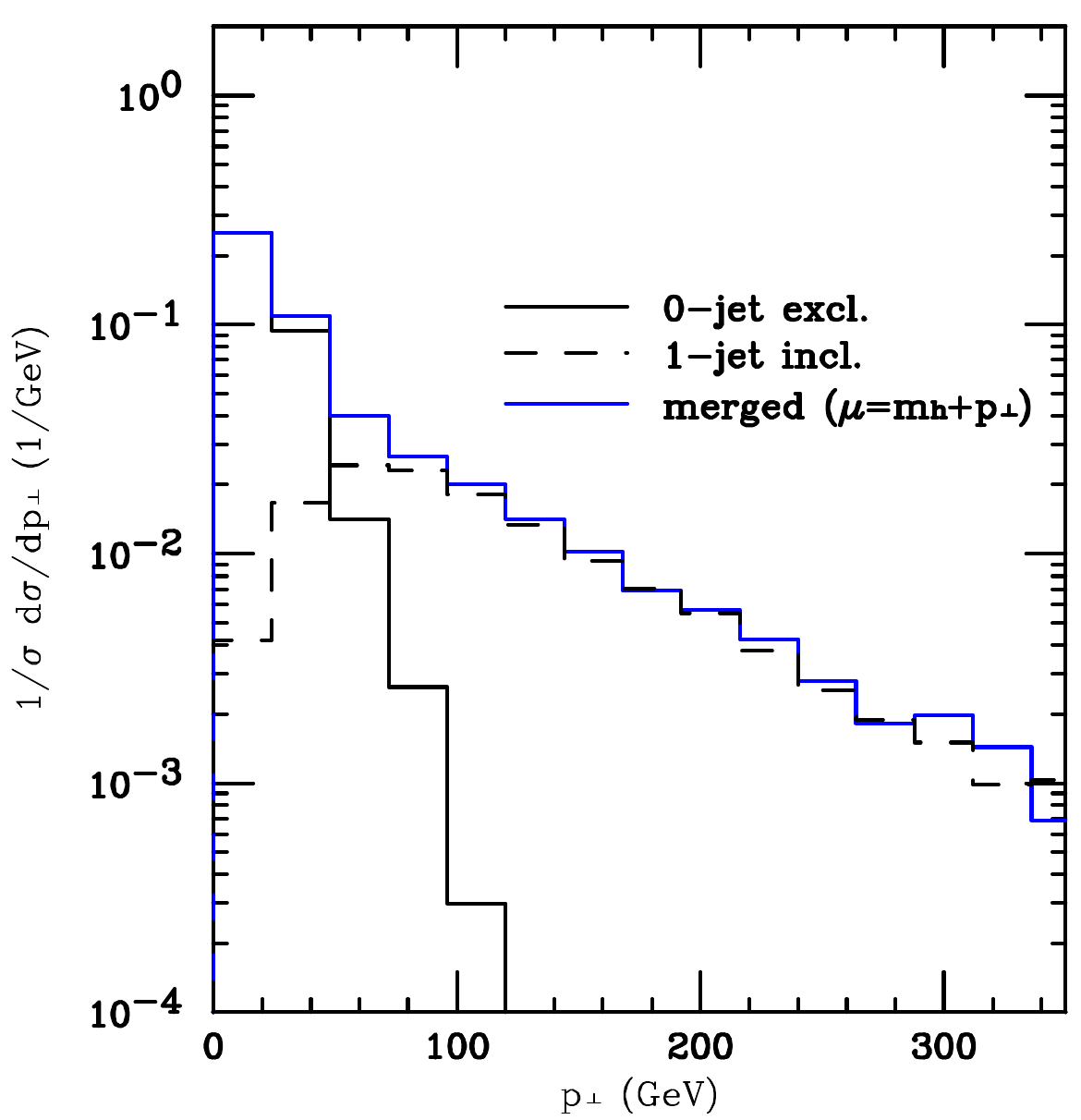}
\hspace{-3mm}
    \includegraphics[width=0.475\linewidth]{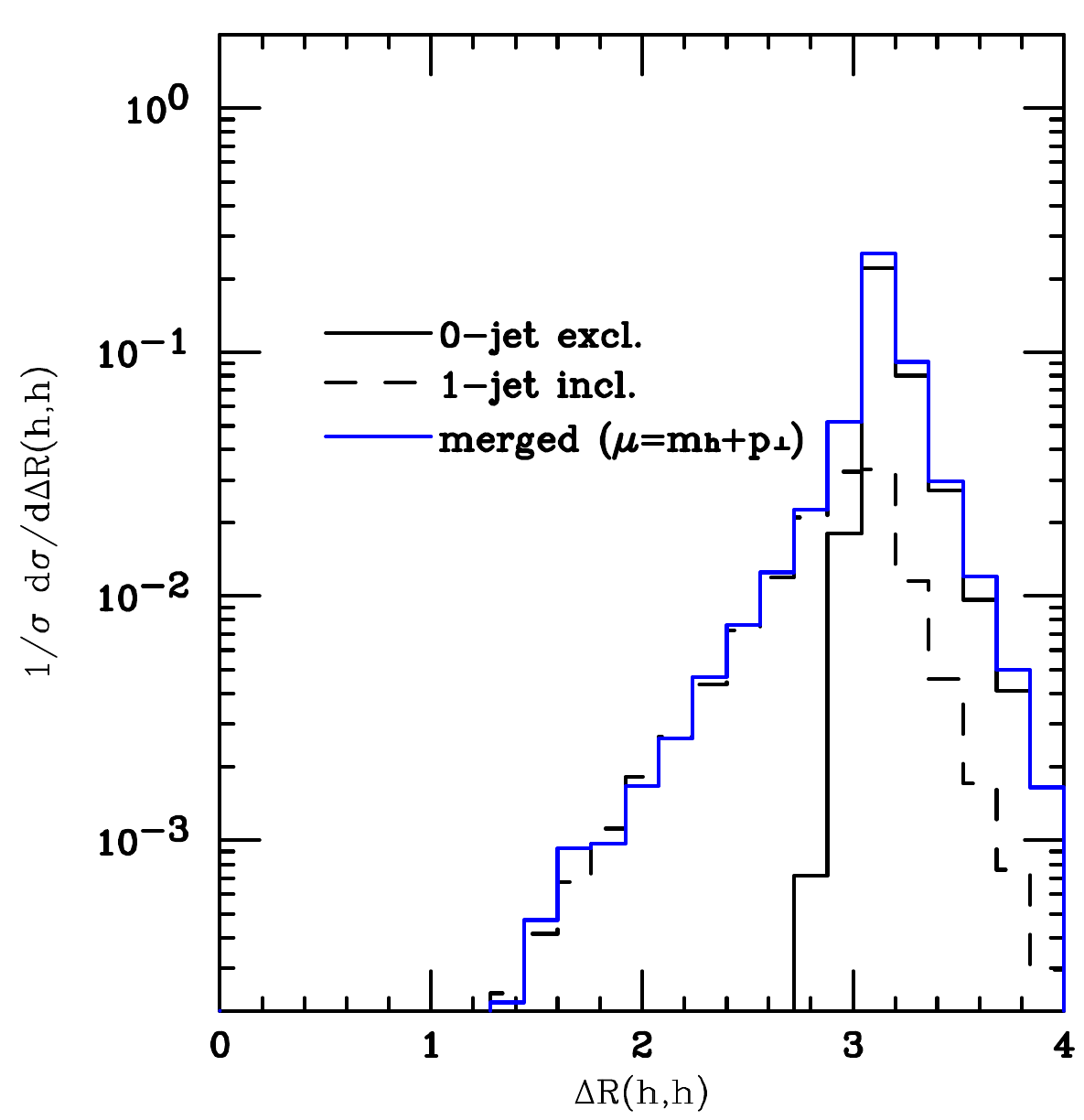}
  \caption{The reconstructed transverse momentum of the Higgs boson
    pair (left) and the distance between the reconstructed Higgs bosons
    (right) resulting from the analysis outlined in the main text broken into their
    individual 0-jet exclusive and 1-jet inclusive contributions. The
    scale was chosen to be $\mu = m_h + p_\perp^{hh}$. }
  \label{fig:pieces}
\end{figure}

We wish to examine the stability of the merged samples against that of
the un-merged samples with respect to scale variations. It is
obvious that sufficiently inclusive quantities should not
differ in a way that will impact the analyses. However, there are quantities for
which the merged sample and the un-merged sample differ substantially. As an
exercise, we examine two such observables here: the distance between the
$(\tau^+\tau^-)$ system and the $(b\bar{b})$ system (equivalent to the
distance between the Higgs bosons), and the transverse momentum of the
$\tau^+\tau^-b\bar{b}$ system (equivalent to the transverse momentum
of the the $hh$ system). Figure~\ref{fig:observables} demonstrates that it is conceivable that both of these observables may
be used for background rejections. Moreover, it is evident that the
largest uncertainties in the $hh$ signal predictions are present in
the exact same region that one would wish to place the cuts in: $p_{T}(hh)
\sim 100$~GeV and $\Delta R (h,h) \sim 3$. What is also important is
the fact that both the 0-jet exclusive and the 1-jet inclusive
signal samples contribute in the region of interest, as demonstrated
in Fig.~\ref{fig:pieces} for the case where $\mu = m_h +
p^{hh}_\perp$. 

\begin{figure}
\centering
    \includegraphics[width=0.45\linewidth]{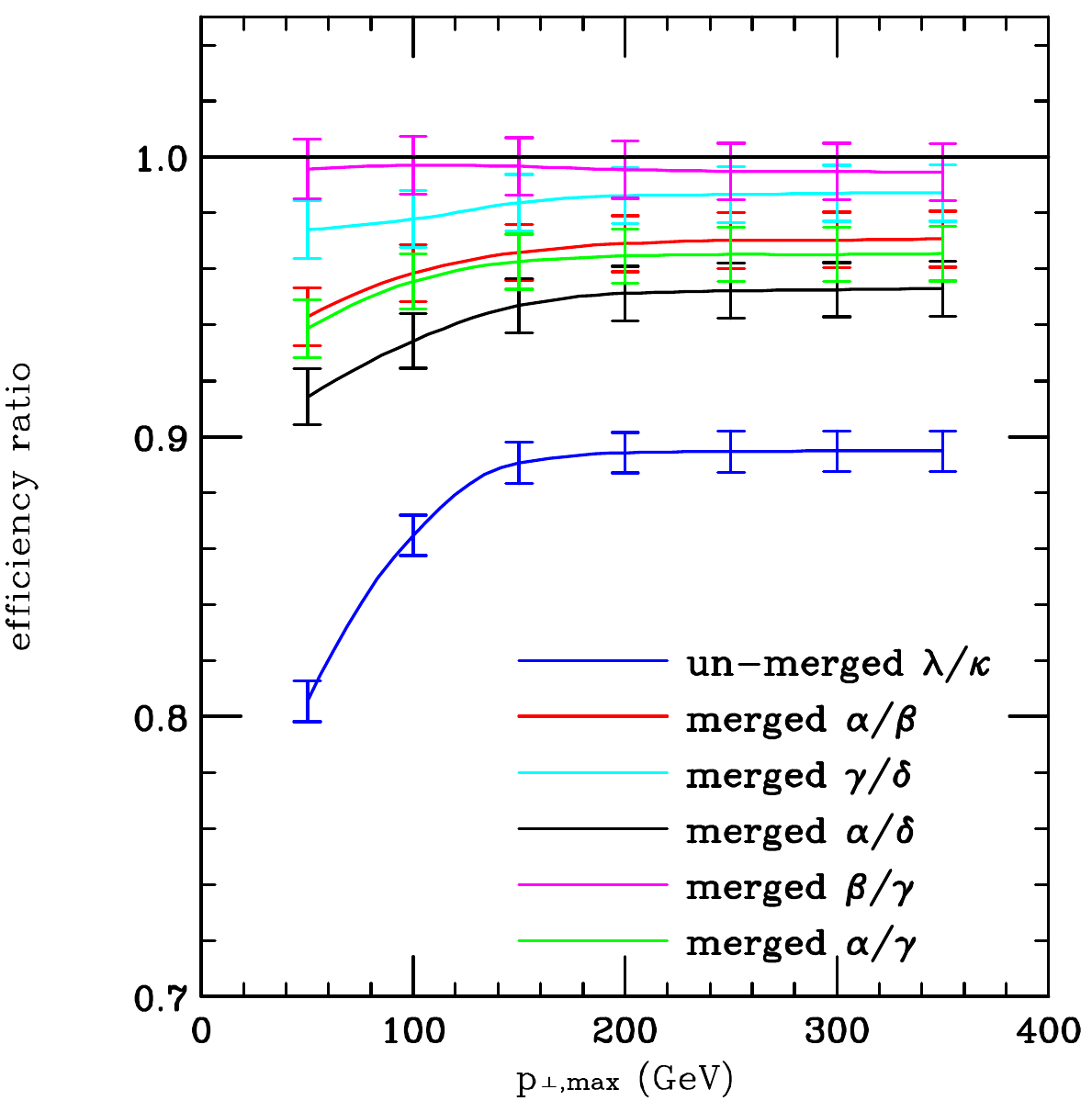}
\hspace{2mm}
    \includegraphics[width=0.45\linewidth]{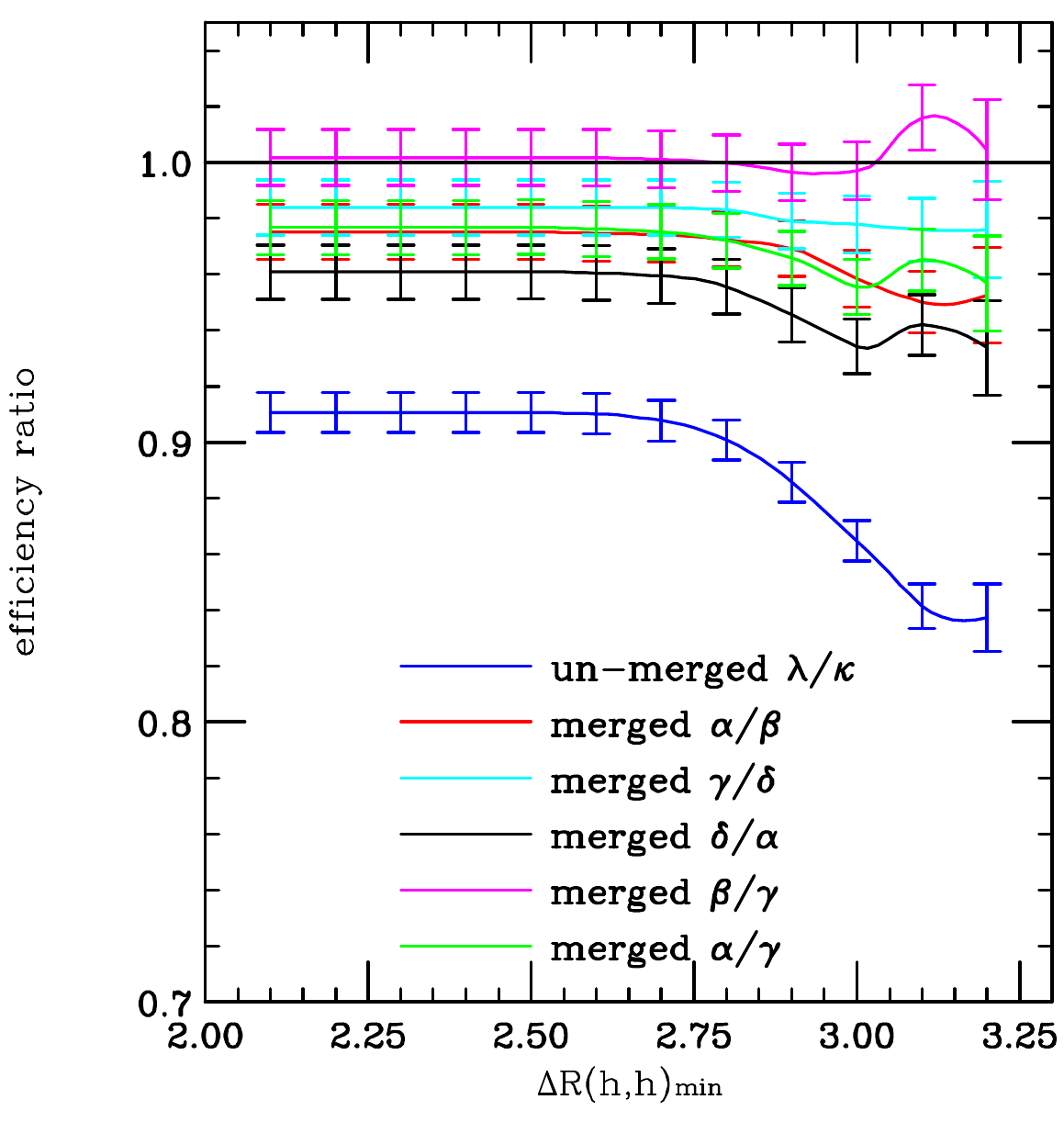}
  \caption{The variation of the ratio of efficiencies with different
    values of the cuts $p_{\perp,\mathrm{max}}$ (left) and $\Delta R
    _{\mathrm{min}} (h,h)$ (right) between two different samples for
    merged and un-merged samples. The sample parameters are: un-merged: $\kappa$: $\mu = m_h$, $\lambda$: $\mu = 2 m_h$.
merged with $\epsilon_{clus}=30$~GeV: $\alpha$: ($\mu=m_h+p^{hh}_\perp$,
$E_{Tclus} = 50$~GeV), $\beta$: ($\mu=2(m_h+p^{hh}_\perp)$,
$E_{Tclus} = 50$~GeV), $\gamma$: ($\mu=m_h+p^{hh}_\perp$,
$E_{Tclus} = 70$~GeV), $\delta$: ($\mu=2(m_h+p^{hh}_\perp)$,
$E_{Tclus} = 70$~GeV), all with $\epsilon_{clus} = 30$~GeV.}
  \label{fig:stability}
\end{figure}

We can further quantify the effect observed in
Fig.~\ref{fig:observables}. The stability of the samples can be
assessed by taking ratios of the efficiencies for different cut
values. If the efficiency does not vary substantially with the cuts, then
the sample (merged or un-merged) is deemed to be stable, and the
theoretical systematic uncertainty on the efficiency of the cut can be
considered to be low. Figure~\ref{fig:stability} shows the variation of the ratio of efficiencies
for different parameters for the merged and un-merged samples:
\begin{eqnarray}
R &=&\mathrm{(cut~efficiency,~sample~i)}/\mathrm{(cut~efficiency,~sample~j)},
\end{eqnarray} 
for cuts on the aforementioned observables which we abbreviate
as $p_{\perp,\mathrm{max}}$ and $\Delta R _{\mathrm{min}}(h,h)$. For
details of the parameters used for each of the samples $\{\alpha, \beta, \gamma,
\delta, \iota, \kappa\}$, see the caption of Fig.~\ref{fig:stability}. The
un-merged samples \{$\kappa$, $\lambda$\} exhibit a fairly substantial change in the ratio of
efficiencies for the two chosen scales, starting from $10\%$ and going up to $\sim 20\%$ for some values of the cuts. This change can be interpreted as a theoretical systematic uncertainty on the efficiency
itself. The merged samples \{$\alpha$, $\beta$, $\gamma$, $\delta$\} perform better, with lower overall
variation of the efficiency ratio, with the deviations always $< 10\%$ as demonstrated in the Fig.~\ref{fig:stability}, while, more
importantly, on average possessing an efficiency variation of $\sim 5\%$. The differences are due to the fact that the chosen observables are sensitive to the behaviour of the extra radiation, which, at high transverse
momentum or large separations between the Higgs bosons, is not
predicted reliably by the parton shower.\footnote{The ratio differs
  from unity since we expect differences in response to cuts in other
  observables between the two samples. This can also be seen in the cut
  flows of Table~\ref{tb:cuts}. Nevertheless, the merged sample ratios
  are still closer to unity overall than those of the un-merged samples,
  demonstrating further the increased accuracy of the calculation.}
\\
For completeness, in Table~\ref{tb:cuts} we show a set of cuts and
resulting cross sections resulting from the analysis. We provide also
explicit cuts on the variables we have examined: $\Delta R (h,h) > 2.8$ and $p^{hh}_\perp<
80$~GeV. The final result for this basic analysis is optimistic, with $S/B \sim
0.4-0.5$ for all samples, leading to a reasonable significance at
600~fb$^{-1}$ of integrated luminosity, expected to be collected at
the LHC in the full phase of Run II. It is worthwhile to note that the
final cross section prediction for the merged samples
($\alpha$ and $\beta$) after all cuts only exhibits a $\sim 2\%$
variation compared to the un-merged samples ($\kappa$ and $\lambda$) which exhibit a $\sim 13\%$ variation.
\begin{table}[t!]
\small
  \begin{center}
    \begin{tabular}{|c|c|c|c|c|c||c|c|c|c|c|}
      \hline
      Process & $\kappa$ & $\lambda$ & $\alpha$&$\beta$
      & $t\bar{t}$ &  $S/B (\kappa)$ & $S/B (\lambda)$  & $S/B (\alpha)$ & $S/B (\beta)$ \\ \hline
      $\sigma$ [fb] & 40.20 & 40.20 & 40.20 & 40.20 & $9\times10^5$ &
      .00004  & .00004 & .00004  & .00004 \\ \hline
      BRs & 2.97 & 2.97 & 2.97& 2.97 & $11000$ & .00027 & .00027 &  .00027 & .00027\\ \hline
      $\tau$ cuts & 0.78 & 0.82 & 0.79  & 0.80  & 296.4 &  .00263 &  .00277 & .00266 &
      .00270 \\ \hline
      fat jet cuts & 0.106 & 0.104 & 0.11 & 0.11 & 0.93 & 0.11 & 0.11 & 0.12 & 0.12
      \\ \hline \hline
      $\Delta R(h,h)$   & 0.106 & 0.100  & 0.099 & 0.101 &
      0.310 & 0.34 & 0.32 & 0.32 & 0.33 \\ \hline
      $p_{\perp}^{hh}$ & 0.103 & 0.089 &  0.095 & 0.093 &  0.207 &
      0.50 
      & 0.43 &0.46 & 0.45  \\ \hline
    \end{tabular}
  \end{center}
  \caption{Cross sections for the $hh$ signal and $t\bar{t}$ \texttt{aMC@NLO} background after series
    of cuts. The un-merged samples $\kappa$ and $\lambda$ have $\mu =
    m_h$ and $\mu = 2 m_h$ respectively and the merged signal samples `$\alpha$' and `$\beta$' have $\mu = m_h +
    p^{hh}_\perp$ and $\mu = 2 (m_h + p^{hh}_\perp)$ respectively, as well as
    $E_{Tclus} = 50$~GeV and $\epsilon_{clus} = 30$~GeV. The final two
    cuts were chosen to be $\Delta R (h,h) > 2.8$ and $p^{hh}_\perp< 80$~GeV.}
\label{tb:cuts}
\end{table}

\section{Conclusions}\label{sec:conclusions}
We have described the implementation of Higgs
  boson pair production merged to the one-jet matrix elements, generated
  using \texttt{OpenLoops}, in the \texttt{HERWIG++}
  event generator. We have examined the systematic uncertainties
  associated with the merging. Moreover, we have provided examples of the magnitude
  of the effects of using the merged samples in a realistic
  analysis. As was demonstrated in this analysis, using the leading order matrix
  elements in conjunction with the parton shower can potentially introduce
  $\mathcal{O}(20\%)$ systematic uncertainties in the predictions of
  the efficiencies of experimental cuts. The uncertainty will
  inexorably propagate to measurements of the Higgs boson
  self-coupling. The merged samples demonstrate theoretical
  uncertainties on the efficiencies that are 10\% or better for the
  examined observables. We expect such conclusions to remain valid for a
  future NLO simulation matched to the parton shower. We thus recommend the use of
  samples that include the merged exact one-jet matrix elements in all
  future phenomenological or experimental analyses of the
  process. The Monte Carlo event generator developed for this project is
  available as an add-on to the \texttt{HERWIG++} event generator at \url{http://www.itp.uzh.ch/~andreasp/hh}.
\appendix
\acknowledgments
We would like to thank Thomas Gehrmann, Massimiliano Grazzini, Stefano
Pozzorini, Jos\'e Zurita and Paolo Torrielli for constructive discussions
during the project. This research is supported in part by the Swiss
National Science Foundation (SNF) under contract 200020-149517 and by
the European Commission through the ``LHCPhenoNet'' Initial Training
Network PITN-GA-2010-264564. 
\bibliography{hhmerge.bib}
\bibliographystyle{JHEP.bst}

\end{document}